\documentclass[fleqn,usenatbib]{mnras}
\usepackage[titletoc,toc]{appendix}
\usepackage{graphicx}
\usepackage{epsfig,float}
\usepackage{multirow}
\usepackage{url}
 \usepackage{color}
\usepackage{amsmath}
\usepackage{amssymb}
\usepackage{epstopdf}
\usepackage{xspace}

\usepackage[colorinlistoftodos]{todonotes}

\newcommand\gbR[1]{}

\newcommand\DGR[1]{}

\newcommand\WGHR[1]{}

\newcommand\BHR[1]{}

\newcommand\red[1]{#1}
\newcommand\magenta[1]{#1}

\newcommand\RR[1]{\textcolor{black}{#1}}

\newcommand\redmagic{\textsc{redMaGiC}\xspace}
\newcommand\gold{\textsc{Gold}\xspace}
\newcommand\NewBPZ{\textsc{BPZ$^*$}}
\newcommand{\imshape}{{\textsc{im3shape}}}
\newcommand{\ngmix}{\textsc{ngmix}}
\newcommand{\metacal}{\textsc{metacalibration}}
\newcommand\eqq[1]{Equation~(\ref{#1})}

\makeatletter
\def\blfootnote{\xdef\@thefnmark{}\@footnotetext}
\makeatother

\usepackage{eso-pic}

\AddToShipoutPictureBG*{%
  \AtPageUpperLeft{%
    \hspace{0.75\paperwidth}%
    \raisebox{-3.5\baselineskip}{%
      \makebox[0pt][l]{\textnormal{DES 2017-0260}}
}}}%
\AddToShipoutPictureBG*{%
  \AtPageUpperLeft{%
    \hspace{0.75\paperwidth}%
    \raisebox{-4.5\baselineskip}{%
      \makebox[0pt][l]{\textnormal{Fermilab PUB-17-293-AE}}
}}}%

\title[Dark Energy Survey Year 1 Results: Redshift distributions of the weak lensing source galaxies]{Dark Energy Survey Year 1 Results: \\
Redshift distributions of the weak lensing source galaxies}
\vspace{-10pt}
\author[DES Collaboration]{
\parbox{\textwidth}{\large
B.~Hoyle$^{1,2\star}$,
D.~Gruen$^{3,4\dagger}$,
G.~M.~Bernstein$^{5}$,
M.~M.~Rau$^{1}$,
J.~De Vicente$^{6}$,
W.~G.~Hartley$^{7,8}$,
E.~Gaztanaga$^{9}$,
J.~DeRose$^{10,3}$,
M.~A.~Troxel$^{11,12}$,
C.~Davis$^{3}$
% ...
\begin{center} (DES Collaboration) \\
\vspace{-40pt}
\end{center} }
}
\begin{document}

\maketitle

\begin{abstract}
We describe the derivation and validation of redshift distribution
estimates \red{and their uncertainties} for the populations of
galaxies used as weak lensing sources in the Dark Energy
Survey (DES) Year 1 cosmological analyses.  The Bayesian Photometric Redshift
(BPZ) code is used to assign galaxies to four redshift bins between $z\approx0.2$ and $\approx1.3$, and to
produce initial estimates of the lensing-weighted redshift
distributions $n^i_{\rm PZ}(z)\propto\mathrm{d}n^i/\mathrm{d}z$ for members of  bin $i$.  \red{Accurate
determination of cosmological parameters depends critically on knowledge of $n^i$ but is
insensitive to bin assignments or redshift errors for individual galaxies.}
The cosmological analyses allow \red{for shifts $n^i(z)=n^i_{\rm
    PZ}(z-\Delta z^i)$ to correct the mean redshift of $n^i(z)$ for biases in $n^i_{\rm PZ}$. The $\Delta z^i$ are
constrained by comparison of independently estimated} 30-band photometric redshifts of galaxies
in the COSMOS field to BPZ estimates made from the DES $griz$ fluxes,
for a sample matched in fluxes, pre-seeing size, and lensing weight to the DES weak-lensing sources.
\red{In companion papers,}
the $\Delta z^i$ of the three lowest redshift bins are further constrained by the angular clustering of
the source galaxies around red galaxies with secure photometric
redshifts at $0.15<z<0.9$.  This paper details the BPZ and COSMOS
procedures, and demonstrates that the cosmological inference is
insensitive to details of the $n^i(z)$ beyond the choice of $\Delta
z^i$.
\red{The clustering and COSMOS validation methods} produce consistent estimates of $\Delta z^i$ in the bins where both can be applied, with
combined uncertainties of $\sigma_{\Delta z^i}=0.015, 0.013, 0.011,$ and $0.022$ in
the four bins.  Repeating the photo-$z$  proceedure instead
using the Directional Neighborhood Fitting (DNF) algorithm, or using the $n^i(z)$ estimated from the matched sample in COSMOS,
yields no discernible difference in cosmological inferences.
\end{abstract}

\blfootnote{Affiliations are listed at the end of the paper.}
\blfootnote{$\star$ corresponding author: \href{mailto:hoyleb@usm.uni-muenchen.de}{hoyleb@usm.uni-muenchen.de}}
\blfootnote{$\dagger$ corresponding author: \href{mailto:dgruen@stanford.edu}{dgruen@stanford.edu}; Einstein fellow}


\clearpage

\section{Introduction}
\label{intro}

The Dark Energy Survey (DES) Year 1 (Y1) data places strong
constraints on cosmological parameters \citep{keypaper} by comparing theoretical models
to measurements of (1) the auto-correlation of the positions of
luminous red galaxies at $0.15<z<0.9$ \citep{wthetapaper} selected by the \redmagic\
algorithm \citep{redmagicSV}; (2) the cross-correlations among weak
lensing shear fields \citep{shearcorr} inferred from the measured shapes of ``source'' galaxies
divided into four redshift bins \citep{shearcat}; and (3) the
cross-correlations of source galaxy shapes around the \redmagic\
(``lens'') galaxy positions \citep{gglpaper}.  There are 650,000 galaxies in the
\redmagic catalog covering the 1321~deg$^2$ DES Y1 analysis area,
and 26~million sources in the primary weak lensing catalog.
For both the lens and the source populations, we rely on DES
photometry in the $griz$ bands\footnote{While there is $Y$ band data available, due to its lower depth, strong wavelength overlap with $z$, and incomplete coverage, we did not use it for photo-$z$ estimation.} to assign galaxies to a redshift bin $i$.
Then we must determine the normalized distribution $n^i(z)$ of galaxies in
each bin.  
This paper describes how the binning and $n^i(z)$
determination are done for the source galaxies. These redshift distributions
are fundamental to the theoretical predictions of the observable lensing
signals.  Uncertainties in the $n^i(z)$ must be propagated into the
cosmological inferences, and should be small enough that induced
uncertainties are subdominant to other experimental uncertainties.  The bin assignments of
the source galaxies can induce selection biases on the shear
measurement, so we further discuss in this paper how this selection
bias is estimated for our primary shear measurement pipeline.  The
assignment of redshifts to the lens galaxies, and validation of the
resultant lens $n^i(z)$'s, are described elsewhere
\citep{redmagicSV,wthetapaper,redmagicpz}. 

\gbR{Insert your favorite photo-z citations next to Sanchez below...}

A multitude of techniques have been developed for estimation of
redshifts from broadband fluxes \citep[e.g.][]{LePhare1,benitez,2001defi.conf...96B,2004PASP..116..345C,2006MNRAS.372..565F,LePhare2,2010A&A...523A..31H,tpz,photozSV,2015MNRAS.452.3710R,2016A&C....16...34H,sadeh16,jvicente}. 
These vary in their statistical methodologies and in their relative reliance on
physically motivated  assumptions vs empirical ``training'' data.
The DES Y1 analyses begin with a photometric redshift algorithm that
produces both a point estimate---used for bin assignment---and an
estimate $p^{\rm PZ}(z)$ of the posterior probability of the redshift of a
galaxy given its fluxes---used for construction of the bins'
$n^i(z).$

The key challenge to use of photo-$z$'s in cosmological inference is the \emph{validation}
of the $n^i(z)$, i.e. the assignment of meaningful error
distributions to them.
The most straightforward method, ``direct''
spectroscopic validation, is to obtain reliable spectroscopic redshifts
for a representative subsample of the sources in each bin. \magenta{Most previous efforts at constraining redshift distributions for cosmic shear analyses used spectroscopic redshifts either as the primary validation method, or to derive the redshift distribution itself \citep{Benjamin13, Jee13, Schmidt13, Bonnett2016, Hildebrandt17}}.
Direct spectroscopic validation cannot, however, currently reach the desired
accuracy for deep and wide surveys like the Y1 DES, because the completeness of
existing spectroscopic surveys is low at the faint end of the DES
source-galaxy distribution \citep{Bonnett2016,Gruen2017},
and strongly dependent on quantities not observed by DES (Hartley et al. in preparation). \RR{In detail the larger area of the DES Y1 analysis compared to other weak lensing surveys, including the DES SV analysis \citep{Bonnett2016}, reduces the statistical uncertainties such that the systematic uncertainties from performing a direct calibration using spectra become dominant.}

The validation for DES Y1 source galaxies therefore uses
high-precision redshift estimates from 30-band photometry of the
COSMOS survey field \citep{Laigle}, which are essentially complete
over the color-magnitude space of the Y1 source catalog, \magenta{in a more sophisticated version of the approach used in \cite{Bonnett2016}}.  This direct
approach is then combined with constraints on $n^i(z)$ derived from
cross-correlation of the source galaxy positions with the \redmagic\
galaxy positions as an
independent method of photometric redshift validation (see, e.g. \citealt{newman} for an introduction to the method and \citealt{xcorrtechnique,redmagicpz,xcorr} for the application to DES Y1). The
cross-correlation redshift technique will be referred to as ``WZ,'' and the
validation based on the 30-band COSMOS photometric redshifts will be
referred to as ``COSMOS,''  and the estimates returned from photo-$z$
algorithms run on
the DES $griz$ photometry will be marked as ``PZ.'' Indeed we suggest reading this paper in conjunction with those of \citealt{xcorrtechnique,xcorr}, which are dedicated to documenting the WZ procedure in greater detail. We also summarise the salient parts of these papers throughout this manuscript and discuss the issue of the failure of the \redmagic\ sample to span the full redshift range of the Y1 lensing sources, which leaves gaps in our knowledge of $n^i$ derived from WZ.

\RR{For the analysis in this work,} the cosmological inference will assume that the
redshift distribution in bin $i$ is given by
\begin{equation}
n^i(z) = n^i_{\rm PZ}(z - \Delta z^i),
\label{Dz}
\end{equation}
where $n_{\rm PZ}^i(z)$ is the distribution returned from the photometric
redshift code using DES $griz$ photometry, and $\Delta z^i$ is a free
parameter to correct any errors resembling a shift of the photo-$z$
result (see also, \citealt{Jee13, Bonnett2016}).  The cosmological inference code is given a probability distribution for
$\Delta z^i,$ which is the normalized product of the probabilities returned by the WZ and
COSMOS analyses.  It is apparent that \eqq{Dz} essentially allows the
mean source redshift returned by the PZ method to be altered by the
information provided by the COSMOS and WZ validation procedures, but the
shape of $n^i(z)$ about its mean retains its PZ determination.

This paper begins in \S\ref{data} with a description of the input
catalogues, real and simulated, for the source redshift inferences and
validation.
\S\ref{algo} describes the photometric redshift algorithms applied to
the DES broadband fluxes.  We
describe the direct COSMOS validation method in \S\ref{cosmos}. The
derivation of WZ constraints from angular clustering is the subject of
\citet{xcorrtechnique}, \citet{redmagicpz}, and \citet{xcorr}.  In \S\ref{res} we
combine these WZ constraints on $\Delta z^i$ with those from COSMOS to
yield the final constraints.
We describe the use of these
redshift constraints as priors for the DES Y1 cosmological inference, including an
examination of the impact of the assumption in \eqq{Dz} and other
known shortcomings in our process, in \S\ref{inference} and conclude in \S\ref{conclusions}.

\red{Aspects of the $n^i(z)$ estimation and validation procedure not immediately required for Y1 lensing analyses will be described in Hoyle et al. (in preparation) and Rau et al. (in preparation).} 

\section{Input catalogs}
\label{data}
 Estimation and validation of the binning and $n^i(z)$ functions for the Y1 source
galaxies require input photometry for these galaxies of course, but
also Dark Energy Camera (DECam, \citealt{Flaugher15}) data \citep{photozSV} and external data on the COSMOS field used for validation.
Finally, our validation uses simulations
of the COSMOS catalog to estimate sample-variance uncertainties
induced by the small sky area of this field.  Fluxes and photo-$z$'s
must be estimated for these simulated galaxies.

\subsection{Lensing sources}
The set of galaxies for which bin assignments and $n^i(z)$ estimates
are desired are the weak lensing (WL) sources defined in the Y1 shear
catalogs documented in \citet{shearcat}.  The primary shear catalog
for DES Y1 is produced by the \metacal\ algorithm
\citep{metacal1,metacal2}, and a secondary catalog using \imshape\ \citep{Zuntz2013} is
used as a cross-check.  \red{For both shear catalogs, we use a common
photo-$z$ catalog based on our best measurements of fluxes (the
``MOF'' catalog described
below) to estimate the $n^i(z)$ of each bin (see~\S\ref{binest} for
details). These $n^i(z)$ differ, however, because
\metacal\ and \imshape\ implement
distinct selection criteria and bin assignments.}

The starting point for either shear catalog is
the Y1 \gold catalog of sources reliably detected on the sum
of the $r,i$, and $z$-band DES images \citep{y1gold}.  Detection and initial
photometry are conducted by the \textsc{SExtractor} software \citep{sextractor}.  
Photometric zeropoints are
assigned to each DES exposure using nightly solutions for zeropoints and
extinction coefficients derived from standard-star exposures.
Exposures from non-photometric nights are adjusted to match those
taken in photometric conditions.

As detailed in \citet{y1gold}, the
photometric calibration is brought to greater color uniformity and
adjusted for Galactic extinction by stellar locus regression (SLR, \citealt{Ivezic:2004a,MacDonald:2004a,High:2009a}):
the $i$-band fluxes are adjusted according to the Galactic extinction
implied by the \citet{SFD} dust map with the \citet{ODonnell94} extinction
law.  Then the zeropoints of other bands
are adjusted to force the stellar color-color loci to a common template.

Fluxes used as input to the photo-$z$ programs for both shear catalogs 
are derived using \ngmix\footnote{https://github.com/esheldon/ngmix} \citep{Sheldon2014,Jarvis2016}, which fits a model to the pixel
values of each galaxy in the \gold
catalog.  The \ngmix\ code fits a highly constrained exponential$+$deVaucouleurs model
to each galaxy: the model is convolved with each exposure's
point-spread function (PSF) and compared to pixels from all individual exposures
covering the source.  The fitting is multi-epoch and multi-band:
pixels of all exposures in all bands are fit simultaneously, assuming
common galaxy shape for all bands and a single free flux per band.  The
fitting is also multi-object: groups of overlapping galaxy images are
fit in iterative fashion, with each fit to a given galaxy subtracting
the current estimate of flux from neighbors.  These ``multi-object
fitting'' (MOF) fluxes are used as input to photo-$z$ estimators for
\imshape\ and \metacal\ catalog member galaxies (although we use a 
different flux measurement for bin assignment in the case of \metacal, see below).

The photo-$z$ assigned to a galaxy depends on its measured multi-band fluxes, which will
vary if there is shear applied to the galaxy. So the photo-$z$ bin to which a galaxy is assigned 
might depend on how much it is sheared, leading to a potential selection bias. 
For \imshape, we have confirmed, using realistic image simulations, that these selection biases are small 
(at or below the one per cent level), and have added a term in the systematic uncertainty of the shear calibration 
to account for them \citep[cf. section 7.6.2 of][called \emph{variation of morphology} there]{shearcat}.
\metacal, on the contrary, can estimate and correct selection biases on the
WL shear inference by producing and re-measuring four artificially sheared
renditions of each target galaxy (by $\gamma_1=\pm0.01$ and $\gamma_2=\pm0.01$, where $\gamma_{1,2}$ are the two components of the shear).  
The selection bias correction in \metacal\ requires knowing whether each source
would have been selected and placed in the same bin if it had been sheared.
It is thus necessary for us to run the photo-$z$ estimation software
not only on the original fluxes, but also on fluxes measured
for each of the four artificially sheared renditions of each galaxy. The latter are not available from the MOF pipeline.

For the \metacal\ catalog, we therefore produce an additional set of photo-$z$ estimates based on 
a different flux measurement made with the \metacal\ pipeline. This measurement makes use of a simplified version of the \ngmix\
procedure described above: the model fit to the galaxies is a
PSF-convolved Gaussian, rather than a sum of exponential and
deVaucouleurs components.  These ``Metacal'' fluxes do not subtract
neighbors' flux. In addition to fluxes, \metacal\ also measures pre-seeing
galaxy sizes and galaxy shapes \citep{shearcat}.

There are thus 6 distinct photo-$z$'s for the WL source
galaxies: one produced using the MOF fluxes for galaxies in either of the
\imshape\ or \metacal\ shape catalogs; one produced using Metacal fluxes of the
as-observed sources in the \metacal\ shape catalog; and four produced
using Metacal fluxes of the four artificially sheared renditions of the
sources in the \metacal\ catalog.

\subsection{COSMOS catalog \& DES $griz$}
\subsubsection{DES fluxes}
Our COSMOS validation procedure depends on having $griz$ photometry
and external redshift estimates for objects in the COSMOS field.  This
field was observed by DES and by community programs using DECam.  These observations were combined, cataloged, and
measured using the same DES pipelines as the survey data; we use the
\texttt{Y1A1 D04} catalog produced as part
of the \gold catalogs \citep{y1gold}.  MOF magnitudes and Metacal sizes are
also measured for all entries in this catalog. The COSMOS-field
observations used herein are $\approx1$~mag deeper than the typical Y1
DES data.  This mismatch must 
be kept in mind when using this field for validation \citep{rau}. 

Zeropoints for the COSMOS images are determined using the same SLR
methods used for the Y1 catalog.  The SLR process is subject to
errors that perturb the calibration.  \RR{ We note that the SLR  adjustment  to the zero points is below 0.03 magnitudes for
most of our data \citep[see fig. A.7 of][]{y1gold}. The  adjustment
for Galactic extinction, which is applied as part of the SLR
procedure, is of order 0.05 magnitudes in most regions of the survey.}
\red{In the Y1 data, because it covers a large area with uncorrelated SLR calibrations, there are many
independent realizations of these errors and they will average away in
the mean $n^i(z)$.
We must keep in mind, however, that
the COSMOS data is based on a single realization of SLR errors, and
must therefore allow for the consequent offset of COSMOS photometry
from the Y1 mean (\autoref{sec:err:slr}).}

\subsubsection{Redshift data and cross-matching}
The COSMOS2015 catalog from \citet{Laigle} provides 
photometry in 30 different UV/visible/IR bands, and probability
distribution functions (PDFs) 
$p^{\rm C30}(z)$ for the redshift of each galaxy based on this
photometry using the \textsc{LePhare} template-fitting code
\citep{LePhare1,LePhare2}.   Typical $p^{\rm C30}(z)$ widths for DES
source galaxies are $\approx0.01(1+z),$ far better than
the uncertainties in BPZ estimates based on DES $griz$ photometry.
In \autoref{laigleerrors} we discuss the influence of errors in
$p^{\rm C30}(z)$'s on our $\Delta z^i$ inferences. 

The validation procedure requires assignment of a $p^{\rm C30}(z)$ to
each DES-detectable source in the COSMOS field.  After limiting the
catalogs to their region of overlap, we associate COSMOS2015
objects with DES \gold objects with 1\farcs5 matching radius.  Only
0.3 per cent of DES-detected sources fail to match a COSMOS2015 source, and
most of these are very near mask boundaries around bright stars or
other peculiar locations.  We conclude that ignoring these unmatched sources
causes an insignificant bias in the inferred redshift distribution.
Of the matched galaxies, 0.4 per cent have no $p^{\rm C30}(z)$ provided in
COSMOS2015,  without explanation.  For these we
\red{synthesize a $p^{\rm C30}(z)$ by averaging those of $\approx10$
nearest neighbors in the space of COSMOS2015 \texttt{ZMINCHI2} and $i$-band
magnitude}, where \RR{ZMINCHI2 is the 30-band photometric redshift point prediction corresponding to the the minimum $\chi^2$ fit between fluxes and templates.}

We remove from the sample galaxies whose fluxes or pre-seeing sizes could not be measured by the 
DES pipelines. We note that such objects would be flagged in the
lensing source catalog and removed. A total of 128,563 galaxies with good DES \gold MOF photometry remain in our
final COSMOS sample. 

\magenta{We also use spectroscopic subsamples of this complete sample of galaxies with COSMOS2015 results later to validate our calibration (cf.~\autoref{sec:err:spec}).}


\subsubsection{PDF rescaling}
\label{pdfrescale}
Following a technique similar to \cite{Bordoloi:2010aa}, we rescale the estimated $p^{\rm C30}(z)$'s to make
them more accurately represent true distribution functions of
redshift.  


The method relies on using the subset of COSMOS2015 galaxies with spectroscopic
redshifts from the literature \citep{2007ApJS..172...70L,2009yCat..21720070L}. While this
subset is not representative of the full photometric
sample \citep{Bonnett2016,Gruen2017}, an excess of outliers in true,
spectroscopic redshift relative to $p^{\rm C30}(z)$ \red{is still an indication
that the rate of ``catastrophic failures'' in COSMOS2015 photo-$z$
determinations is higher than that estimated by \citet{Laigle}.  The
procedure described here is not a panacea but will lessen such
discrepancies.}

\begin{figure}
\centering
\includegraphics[width=0.9\linewidth]{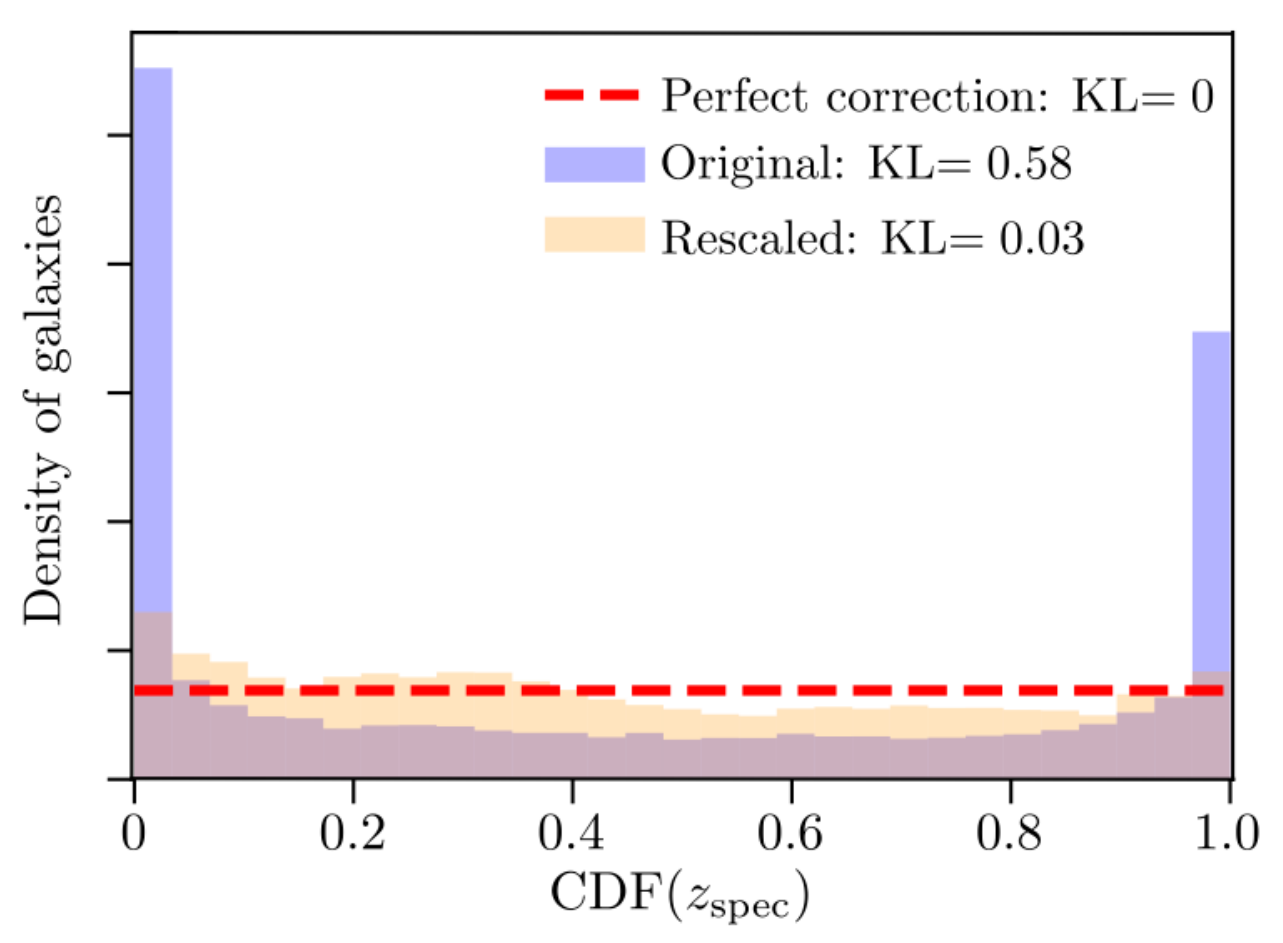}
\caption{\label{RescaledCosmos} The effect of rescaling the COSMOS2015
  photometric redshift PDFs using the  Probability Integral Transform (PIT) distribution. The PIT is the redshift
  cumulative distribution function (CDF) values of the full sample of
  DES-detected sources evaluated at the
  spectroscopic redshift, for those sources with known $z_{\rm spec}$.
   The original PDFs (blue) depart significantly from the expected
   uniform distribution (red dashed line).  The $p^{\rm C30}(z)$ rescaling
   procedure yields the orange histogram, much improved, as confirmed
   by the 
  value of the Kullback-Leibler divergence between the histogram and a
  uniform distribution.}

\end{figure}

For each galaxy in COSMOS2015 having a spectroscopic
redshift and matching a DES detection, $p^{\rm C30}(z)$ is integrated to a cumulative
distribution function (CDF) $0<c(z)<1.$ The value $c(z_{\rm spec})$ for a distribution of objects is the Probability Integral Transform (PIT) \citep{PIT,Angus:1994:PIT:209190.209206}.
If $p^{\rm C30}(z)$ is a true, statistically rigorous PDF of the spectroscopic redshifts, the PIT
values should be uniformly distributed between 0 and 1.  
In Figure~\ref{RescaledCosmos} we show in blue the distribution
of PIT values for the original $p^{\rm C30}(z)$'s.
The peaks at 0 and at 1 indicate that the widths of the $p^{\rm C30}(z)$ are
underestimated and need to be broadened, and the asymmetry means
that a small global offset should be
applied to them.

We recalibrate the $p^{\rm C30}(z)$'s by positing that the true PDF can
be well approximated by applying the following transformation to the original $p^{\rm C30}(z)$:
\begin{equation}
p^{\rm C30}(z) \rightarrow A \cdot p^{\rm C30}(z) \otimes {\mathcal N}(\mu,\sigma) +
\frac{1-A}{\pi \gamma \left[ 1 +
    \left(\frac{z-(z_0+\mu)}{\gamma}\right)^2\right]}.
\end{equation}
In the first term, $p^{\rm C30}(z)$ is slightly broadened and shifted by
convolution with a Gaussian of width $\sigma$ and center $\mu$.  The
second term adds in a Cauchy distribution about the median value $z_0$
of the original $p^{\rm C30}(z)$ to allow for long tails.  The free parameters
$\{A, \mu, \sigma, \gamma\}$ are found using the \texttt{Nelder-Mead}
method of \textbf{scipy.stats.minimise} to minimize  the
Kullback-Leibler (KL)
divergence between the histogram of CRPS values and the expected
uniform distribution.
The best-fitting recalibration parameters are derived using a randomly selected 50\% of the spectroscopic catalog and then validated on the remaining 50\%.  The histogram of CRPS of the validation samples after $p^{\rm C30}(z)$ recalibration is shown in Figure~\ref{RescaledCosmos} by the orange histograms.

Going into further detail: we determine the best-fitting remapping parameters
independently for six subsets in bins of $i$-band MAG\_AUTO \citep{sextractor}
magnitude bounded by [16.1, 20.72, 21.48, 21.98, 22.40, 23.03, 99]. The
bins are chosen so that they are each populated by approximately 4000
spectra. Remapping in bins of magnitude is seen to yield lower KL
values than remapping in redshift bins, or with no binning. We find
that the KL values of the training data and the validation data are
very similar, indicating that we are not over-fitting. The KL
divergences in the first and last bin decrease from $0.88 \rightarrow
0.14$ and $0.52\rightarrow 0.13$, respectively, with even greater
improvement for the full sample as noted in
Figure~\ref{RescaledCosmos}.
The only parameter relevant for the mean redshift calibration performed in \autoref{cosmos} is the shift in the mean of the $p(z)$, $\mu$. The sizes of these in each magnitude bin are all $|\mu|\le0.001$, much smaller than the uncertainty of our ensemble mean redshifts. 
 



\subsection{Simulated sky catalogs}
\label{buzzard}

We also draw upon simulated data sets generated specifically for the
DES collaboration. Specifically, we make use of the
\textsc{Buzzard-v1.1} simulation, a mock DES Y1 survey created from a
set of dark-matter-only simulations. This simulation and the galaxy
catalog construction are described in detail elsewhere
\citep{DeRose2017, Wechsler2017, simspaper}, so here we provide only a
brief overview. \textsc{Buzzard-v1.1} is constructed from a set of 3
$N$-body simulations run using \textsc{L-GADGET2}, a version of
\textsc{GADGET2} modified for memory efficiency, with box lengths
ranging from 1--4~$h^{-1}$Gpc from which light-cones were constructed
on the fly.

Galaxies are added to the simulations using the Adding Density
Dependent GAlaxies to Light-cone Simulations algorithm
[\textsc{ADDGALS}, \citealt{Wechsler2017}]. Spectral energy
distributions (SEDs) are assigned to the galaxies from a training set
of spectroscopic data from SDSS DR7 \citep{Cooper2011} based on local
environmental density. These SEDs are integrated in the DES pass
bands to generate $griz$ magnitudes.  Galaxy sizes and ellipticities
are drawn from distributions fit to SuprimeCam $i^{'}$-band data \citep{spcam}. The
galaxy positions, shapes and magnitudes are then lensed using the the
multiple-plane ray-tracing code, Curved-sky grAvitational Lensing for
Cosmological Light conE simulatioNS [CALCLENS,
\citet{Becker2013}]. The simulation is cut to the DES Y1 footprint,
and photometric errors are applied to the lensed magnitudes by copying
the noise map of the \texttt{FLUX\_AUTO} measurements in the real
catalog. More explicitly, the error on the observed flux is determined only
by the limiting magnitude at the position of the galaxy, the exposure
time, and the noise-free apparent magnitude of the galaxy itself.

\subsubsection{Science sample selection in simulations}
The source-galaxy samples in simulations are selected so as to roughly
mimic the selections and the redshift distributions of the \metacal\
shear catalog described in \cite{shearcat}. This is done by first
applying flux and size cuts to the simulated galaxies so as to mimic
the thresholds used in the Y1 data by using the Y1 depth and PSF
maps. The weak lensing effective number density $n_{\rm eff}$ 
  in the simulation \red{is matched to a preliminary version of
  the shape catalogs, and is about 7 per cent higher than for the final,
  unblinded \metacal\ catalog.}  Truth values for redshift, flux and
shear are of course available as well as the simulated measurements.

\gbR{Joe: question from Ashley, can you be more precise than 10--20\%? I am guessing it depends on choices of bins etc., so you cannot.}
\DGR{effective number density is 5.4 vs 5.07 in the data, described that above}

COSMOS-like catalogs are also generated from the Buzzard simulated
galaxy catalogs
by cutting out 367 non-overlapping COSMOS-shaped footprints from the simulation.

\section{Photometric redshift estimation}
\label{algo}

\RR{
In this section we describe the process of obtaining photometric redshifts for DES galaxies.  We note that we only use the $g,r,i,z$ DES bands in this process. We have found that the $Y$ band adds little to no predictive power.
}
\subsection{Bayesian Photometric Redshifts (BPZ)}
\label{bpz}


Posterior probabilities $p^{\rm PZ}(z)$ were calculated for each source galaxy
using \NewBPZ, which is a variant of the Bayesian algorithm described
by \citet{benitez}, and has been modified to provide the photometric
redshift point predictions and PDFs required by the DES collaboration
directly from \textsc{fits}-format input fluxes, without intermediate
steps.  The \NewBPZ\ code is a distilled version of the distributed
\textsc{BPZ} code, and in particular assumes the synthetic template
files for each filter have already been generated. \red{Henceforth we
  will refer to these simply as ``BPZ'' results.}

\subsubsection{Per-galaxy posterior estimation}

The redshift posterior is calculated by marginalising over a set of
interpolated model spectral templates, where the likelihood of a
galaxy's photometry belonging to a given template at a given redshift
is computed via the $\chi^2$ between the observed photometry and those
of the filter passbands integrated over the model template. \magenta{The model templates are grouped into three classes, nominally to represent elliptical, spiral and star-burst galaxies. These classes, it is assumed, follow distinct redshift-evolving luminosity functions which can be used to create a magnitude-dependent prior on the redshift posterior of each object, a.k.a. the ``luminosity prior''. The prior comprises two components, a spectral class prior which is dependent only on observed magnitude, and the redshift prior of each class---which is itself also magnitude dependent (see \citealt{benitez} for more detail)}.


\magenta{Six base} template spectra for BPZ are generated based on original models by
\cite{Coleman80} and \cite{Kinney96}. The stellar locus regression
used for the DES Y1 data ensures uniformity of color across the
footprint, but there may be small differences in calibration with
respect to the empirical templates we wish to use. 
Moreover, these
original templates are derived from galaxies at redshift zero, while
our source galaxies cover a wide range in redshift, with an
appreciable tail as high as $z\sim1.5$. The colors of galaxies evolve significantly over this redshift range, even
at fixed spectral type. Failure to account for this evolution can
easily introduce biases in the redshift posteriors that subsequently
require large model bias corrections (see \citealt{Bonnett2016}, for
instance). To address these two issues, we compute
evolution/calibration corrections to the template fluxes.

We match low-resolution spectroscopic redshifts from the PRIMUS DR1 dataset
\citep{Coil11,Cool13} to high signal-to-noise DES photometry and
obtain the best fit of the six basic templates to each of the highest
quality PRIMUS objects (quality = 4) at their spectroscopic
redshift. The flux of each template in each filter is then corrected
as a function of redshift by the median offset between the DES
photometry and the template prediction, in a sliding redshift window
of width $\delta z=0.06$. The calibration sample numbers 72,176 galaxies and
reaches the full depth of our science sample ($i_{{\rm DES}}<23.5$) while
maintaining a low rate of mis-assigned redshifts.\footnote{The outlier
  fraction of $7.85\%$ quoted in \cite{Cool13} includes all objects
  that lie more than $\delta z>0.025$ from their true redshift. The
  difference in template photometry caused by such a small change in
  redshift is well within the scatter of our computed DES - template
  offsets. Of greater concern is the fraction of objects with large
  redshift differences, which is $<4\%$.}
Although the incompleteness in
PRIMUS is broadly independent of galaxy color \citep{Cool13} and each
template is calibrated separately, we nevertheless expect small
residual inaccuracies in our calibration to remain. 
Our COSMOS and WZ validation strategies serve to calibrate such errors
in BPZ assignments.

A complete galaxy sample is required for deriving the luminosity prior we use with BPZ. No spectroscopic samples are complete to the limit of our source galaxy sample, and so we turn to the accurate photometric redshift sample in the COSMOS field from \cite{Laigle}, which is complete to the depth of our main survey area despite being selected in the $K$-band. The prior takes the form of smooth exponential functions (see \citealt{benitez}), which we fit to the COSMOS galaxy population by determining galaxy types at their photometric redshift. Because BPZ uses smooth functions rather than the population directly, the luminosity prior used for obtaining posterior redshift probabilities does not replicate the high-frequency line-of-sight structure in the COSMOS field.

BPZ is run on the MOF fluxes (see \S\ref{data}) to determine $p^{\rm PZ}(z)$ for \metacal\ and \imshape, while for the five \metacal\ catalogs---the real one and
the four artificially sheared versions---BPZ is run on the \metacal\
fluxes to determine bin assignments (cf.~\S\ref{binest} for details). The luminosity prior is constructed from MOF $i$-band fluxes for
both catalogues. For the Buzzard simulated galaxy catalogs, BPZ is run
  on the single mock flux measurement produced in the simulation.

\RR{We also explored a further post-processing step as in \S\ref{pdfrescale}, but applied to the DES BPZ photo-z PDFs. We used the spectroscopic training data, which is not used in BPZ, to recalibrate the PDFs in bins of $i$-band magnitude. We find that this rescaling did not noticeably change the mean or widths of the PDFs on average, and that the statistical properties of the redshift distributions in each tomographic bin also remain unchanged.}

\subsubsection{Known errors}
\label{knownerrors}
During BPZ processing of the Y1 data, three configuration and software errors were
made.  

First, the \metacal\ catalogs were processed using MOF $i$-band magnitudes for evaluating the BPZ prior
rather than Metacal fluxes. This is internally consistent for BPZ,
but the use of flux measurements that do not exist for artificially sheared galaxies means
that the \metacal\ shear estimates are not properly corrected for
selection biases resulting from redshift bin assignment.  We note that small perturbations to the flux used for assigning the luminosity prior have very little impact on the resulting mean redshift and the \emph{colors} used by BPZ in this run are correctly measured by \metacal\ on unsheared and sheared galaxy images. Rerunning
BPZ with the correct, Metacal inputs for $i$-band magnitude on a subset of galaxies indicates that the
induced \RR{multiplicative} shear bias is below $0.002$ in all redshift bins, well below
both the level of statistical errors in DES Y1 and our uncertainty in shear bias calibration. 
We therefore decide to tolerate the resulting systematic uncertainty.

Second, the SLR  \RR{adjustments}  to photometric zeropoints were not applied
to the observed Metacal fluxes in the Y1 catalogs before input to BPZ.  The
principal result of this error is that the observed magnitudes are no
longer corrected for Galactic extinction.  This results in a shift in
the average $n^i(z)$ of the source population of each bin, and a
spatially coherent modulation of the bin occupations and redshift
distributions across the survey footprint.  In \S\ref{cosmos} we
describe a process whereby the mean $n^i(z)$ can be accurately
estimated by mimicking the SLR errors on the COSMOS field.  In
Appendix~\ref{inhomogeneity} we show that the spurious spatial variation of the
redshift distributions causes negligible errors in our estimation of
the shear two-point functions used for cosmological inference, and
zero error in the galaxy-galaxy lensing estimates.

\red{Finally, when rewriting BPZ for a faster version, \NewBPZ, two bugs were introduced in the prior implementation, one causing a bias for bright galaxies ($i$ band magnitude $<18.5$) and another which forced uniform prior abundance for the three galaxy templates. These bugs were discovered too late in the DES Y1 analysis to fix. They cause differences in $\Delta z$ that are subdominant to our calibration uncertainties (below 0.006 among all individual bins). In addition, they are fully calibrated by both COSMOS (which uses the same implementation) and WZ, and hence do not affect our cosmological analysis.} \RR{We have since implemented all of the above bug fixes, and applied the SLR  \RR{adjustments}  correctly, and find negligible changes in the shape and mean of the BPZ PDFs, which are fully within the combined systematic uncertainties.}

\DGR{This misses another known error: the bugs in BPZ$^{\star}$.}

\subsubsection{Per-galaxy photo-$z$ precision}

\begin{figure}
\centering
\includegraphics[scale=0.55,clip=true,trim=10 10 10 15]{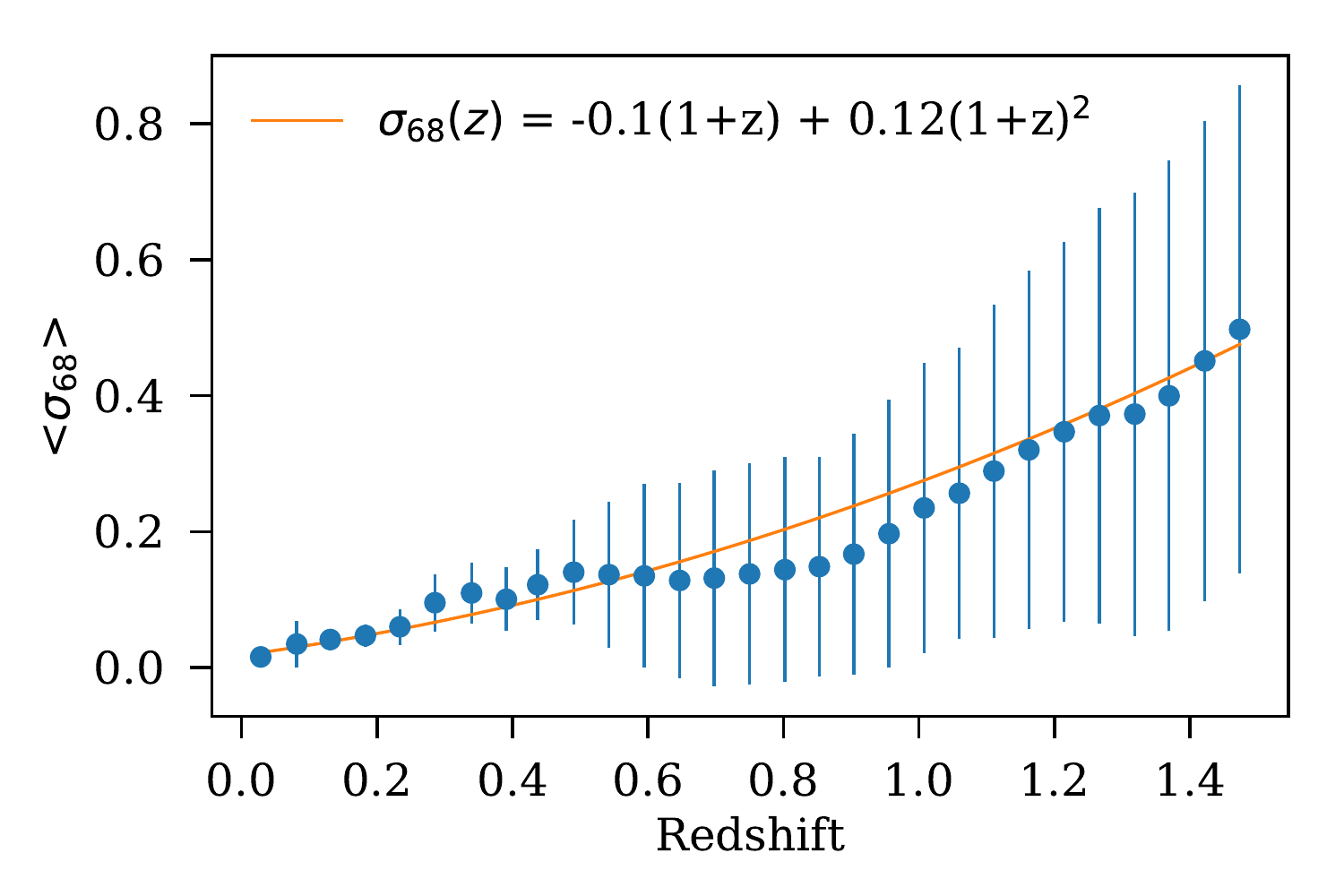}
\caption{\label{sigma_bpz} The average width of the posterior
  distributions of BPZ photometric redshifts \RR{for data selected in bins of mean BPZ redshift.} The posterior width is defined as the 68\%
  spread of the PDF $p^{\rm PZ}(z)$ about its median. The error bars correspond to the standard deviation of the individual source's $\sigma_{68}$ around the average.}
\end{figure}

While $n^i(z)$ are the critical inputs to cosmological inference,
it is sometimes of use to know
the typical size of the redshift uncertainty for individual galaxies.
We define $\sigma_{68}$ for each $p^{\rm PZ}(z)$ as the half-width of the \magenta{68 percentile} region around
  the median.  
We select 200,000 galaxies from the \metacal\ catalog at random, and
determine the average $\sigma_{68}$ in bins of redshift according to
the median of $p^{\rm PZ}(z)$.  We find
that this mean $\sigma_{68}(z)$ is well fit by a quadratic polynomial in \RR{mean BPZ} redshift and present
the best-fitting parameters in Figure~\ref{sigma_bpz}.   

Further metrics
of the performance of individual galaxies' photo-$z$'s but with respect to truth redshifts
  are provided in \S\ref{standard_metrics}.


\subsection{Directional Neighborhood Fitting (DNF)}   
Directional Neighborhood Fitting (DNF) \citep{jvicente} is a
machine-learning algorithm \magenta{for galaxy photometric redshift estimation. We have applied it to reconstruct the redshift distributions for the \metacal\ catalogs}. DNF takes as reference a training
sample whose spectroscopic redshifts are known. Based on the training
sample, DNF constructs the prediction hyperplane that best fits the
neighborhood of each target galaxy in \magenta{multiband flux} space. It then uses
this hyperplane to predict the redshift of the target galaxy. The key
feature of DNF is the definition of a new neighborhood, the
Directional Neighborhood. Under this definition --- and leaving apart
degeneracies corresponding to different galaxy types --- two
galaxies are neighbors not only when they are close in the Euclidean
\magenta{multiband flux} space, but also when they have similar
\magenta{relative flux} in different bands, i.e. colors. In this way,
the neighborhood does not extend in multiband flux hyperspheres but in
elongated hypervolumes that better
represent similar color, and presumably similar redshift. 
As described in \S\ref{binest}, these DNF photo-$z$ predictions are used to classify the galaxies in tomographic redshift bins. 

A random sample from the $p^{\rm PZ}(z)$ of an object is approximated in the DNF method by the redshift of the nearest neighbor within the training sample. It is used as the sample for $n^i(z)$ reconstruction and interpreted in section \S\ref{binest} as a random draw from the underlying per-galaxy posterior.

The training sample used for Y1 DNF prediction was collected by the
DES Science Portal team \citep{julia} from different spectroscopic surveys and 
includes the  VIPERS 2nd data release \citep[][]{2016arXiv161107048S}.
The validation of the predictions was based on COSMOS2015 photo-$z$'s. 
Objects near the COSMOS data were removed from the training sample. 
\magenta{Since the machine learning algorithm can correct for imperfections in the input photometry giving a representative training set, both training and photo-$z$ predictions are based on Metacal photometry without SLR- \RR{adjustments}, for all runs on DNF.}

The fiducial DES Y1 cosmological parameter estimation uses the BPZ
photo-$z$'s, and \citet{keypaper} demonstrate that these estimates are
robust to substitution of DNF for BPZ. 

The $n^i(z)$ distributions of
BPZ and DNF are not expected to be identical, because the algorithms
may make different bin assignments for the same source.  We therefore
do not offer a direct comparison.  We do, however, repeat for DNF all
of the validation processes described herein for the BPZ $n^i(z)$
estimates.  The results for DNF are given in
Appendix~\ref{dnfresults}.

\subsection{Binning and initial $n^i(z)$ estimation}
\label{binest}

\begin{table*}
\caption{Binning, $n^i(z)$ estimation, and mean $z$ calibration for the variants of the shear and photo-z catalogs}
\label{bectable}
\begin{center}
\begin{tabular}{l|lcc|}
\hline
\hline
shear catalog & step & BPZ & DNF  \\
\hline 
& binned by: & Metacal $griz$ $\langle z_j\rangle$ & Metacal $griz$ $\langle z_j\rangle$ \\
\metacal  & $n^i(z)$ by stacking: & MOF $griz$ $z_j^{\rm PZ}$ &  Metacal $griz$ $z_j^{\rm PZ}$ \\ 
 & calibration by: & COSMOS + WZ & COSMOS + WZ \\
 \hline
& binned by: & MOF $griz$ $\langle z_j\rangle$ & --- \\
\imshape  & $n^i(z)$ by stacking: & MOF $griz$ $z_j^{\rm PZ}$ &  --- \\ 
 & calibration by: & COSMOS + WZ & --- \\
 \hline
 \end{tabular}
 \end{center}
 \end{table*}
 

Both photo-$z$ codes yield 6 different posterior distributions $p^{\rm
  PZ}(z_j)$ for each galaxy $j$ in the Y1 shape catalogs, conditional on either the MOF, the unsheared Metacal, or the four sheared Metacal flux measurements. In this section, we describe how these are used to define source redshift bins $i$ and provide an initial estimate of the lensing-weighted $n^i(z)$ of each of these bins. \red{\autoref{bectable} gives an overview of these steps.}

Galaxies are assigned to bins based on the expectation value of their
posterior, $\langle z_j\rangle=\int z_j\,p^{\rm PZ}(z_j) \, \mathrm{d}z_j$. We use four bins between the limits $[0.20,0.43,0.63,0.90,1.30]$. 
\magenta{These tomographic boundaries exclude $\langle z_j\rangle<0.2$ and $\langle z_j\rangle>1.3$ that have large photo-$z$ biases. We place three tomographic bins at $\langle z_j\rangle<0.9$ with approximately equal
effective source density $n_{\mathrm{eff}}$, a proxy for the statistical uncertainty of shear signals in the \metacal\ catalog, since $z=0.9$ is the
upper limit of the WZ constraints. The fourth bin, $0.9<\langle z_j\rangle<1.3,$ is
thus validated only by the COSMOS method.} 

For \metacal\ sources, this bin assignment is made based on the $\langle z_j\rangle$ of the photo-$z$ run on Metacal photometry, instead of MOF photometry. The reason for this is that flux measurements, and therefore photo-$z$ bin assignments, can depend on the shear a galaxy is subject to. This can cause selection biases in shear due to photo-$z$ binning, which can be corrected in \metacal. The latter requires that the bin assignment can be repeated using a photo-$z$ estimate made from measurements made on artificially sheared images of the respective galaxy \citep[cf.][]{metacal1,metacal2,shearcat}, and only the Metacal measurement provides that. 

For \imshape\ sources, the bin assignment is made based on the
$\langle z_j\rangle$ of the photo-$z$ run on MOF photometry, which has
higher $S/N$ and lower susceptibility to blending effects than Metacal
photometry.  This provides more precise (and possibly more accurate) photo-$z$ estimates.

We note that this means that for each combination of shear and photo-$z$ pipeline, bin assignments and effective weights of galaxies are different. The redshift distributions and calibrations derived below can therefore not be directly compared between the different variants.

The stacked redshift distribution $n^i(z)$ of each of the tomographic
bins is estimated by the lensing-weighted stack of random samples
$z_j^{\rm PZ}$ from the $p^{\rm PZ}(z_j)$ of each of all galaxies $j$ in bin
$i$. Given the millions of galaxies in each bin, the noise due to using only one random
sample from each galaxy is negligible. For both the \metacal\ and
\imshape\ catalogs, we use random samples from the $p^{\rm PZ}(z)$
estimated by BPZ run on MOF photometry to construct the $n^i(z)$, 
this being the lower-noise and more reliable flux estimate.
In the case of DNF, we use the Metacal photometry run for both the binning and initial $n^i(z)$ estimation.

By the term \emph{lensing-weighted} above, we mean the effective
weight $w^{\rm eff}_j$ a source $j$ has in the lensing signals we
measure in \citet{shearcorr} and \citet{gglpaper}. In the case of
\metacal, sources are not explicitly weighted in these papers. Since
the ellipticities of galaxies in \metacal\ have different responses to
shear \citep{metacal1,metacal2}, and since we measure correlation
functions of \metacal\ ellipticities that we then correct for the mean
response of the ensemble, however, sources do have an effective weight
that is proportional to their response. As can be derived by
considering a mixture of subsamples at different redshifts and with
different mean response, the correct redshift distribution to use is
therefore one weighted by $w_j^{\rm
  eff}\propto(R_{\gamma_{1,1},j}+R_{\gamma_{2,2},j}),$ where the $R$'s
are shear responses defined in \citet{shearcat}. In the case of \imshape, explicit weights $w_j$ are used in the measurements, and sources have a response to shear $(1+m_j)$ with the calibrated multiplicative shear bias $m_j$ \citep{shearcat}. The correct effective weights for \imshape\ are therefore $w_j^{\rm eff}\propto(1+m_j)\times w_j$.

We note that for other uses of the shape catalogs, such as with the
optimal $\Delta\Sigma$ estimator \citep{Sheldon2004}, the effective
weights of sources could be different, which has to be accounted for
in the photo-$z$ calibration.

\section{Validating the redshift distribution using COSMOS multi-band photometry}
\label{cosmos}

\magenta{In \cite{Bonnett2016} we made use of COSMOS photometric redshifts as an independent estimate 
and validation of the redshift distribution of the weak lensing source galaxies. 
We made cuts in magnitude, FWHM and surface brightness to the source catalogue from DECam images in the COSMOS 
field that were depth-matched to the main survey area. These cuts approximated the selection function of the shape 
catalogues used for the cosmic shear analysis. Similar techniques that find COSMOS samples of galaxies matched 
to a lensing source catalog by a combination of magnitude, color and morphological properties
have been applied by numerous studies \citep{Applegate2014,Hoekstra2015,Okabe2016,Cibirka2017,Amon17}.
In the present work, we modify the approach to reduce statistical and systematic uncertainty on its estimate of mean redshift
and carefully estimate the most significant sources of systematic error.}

We wish to validate the $n^i(z)$ derived for a target sample A of galaxies
using a sample B with known redshifts.  Ideally, for every galaxy in
A, we would find a galaxy in B that looks exactly like it when
observed in the same conditions.  The match would need to be made in
all properties we use to select and weight the galaxy in the weak
lensing sample that also correlate with redshift.

Then the mean redshift distribution of the matched B galaxies, weighted
the same way as the A galaxies are for WL measures, will yield the desired $n^i(z)$.
This goal is unattainable without major observational, image processing and simulation efforts, but we can approximate it with a method related to the one of \citet{Lima2008} and estimate the
remaining uncertainties.  We also need to quantify uncertainties
resulting from the finite size of sample B, and from possible errors
in the ``known'' redshifts of B.  Here our sample A are the galaxies
in either the \imshape\ or \metacal\ Y1 WL catalogs, spread over the footprint of DES Y1, and sample B is
the COSMOS2015 catalog of \citet{Laigle}.

\subsection{Methodology}
\label{cosmosresample}
We begin by selecting a random subsample of 200,000 galaxies from each
WL source catalog, spread over the whole Y1 footprint, and assigning to each a match in the COSMOS2015
catalog.  The match is made by $griz$ MOF flux and pre-seeing size (not by position), and the matching algorithm proceeds as follows, for each galaxy in the WL source sample:
\begin{enumerate}
\item Gaussian noise is added to the DES $griz$ MOF fluxes and sizes of the COSMOS
galaxies until their noise level is equal to that of the target
galaxy.  COSMOS galaxies whose flux noise is above that of the
target galaxy are \red{considered ineligible for matching.  While this
removes 13\% of potential DES-COSMOS pairs, this is unlikely to
induce redshift biases, because the
noise level of the COSMOS $griz$ catalog is well below that of the Y1
survey in most regions of either, so it should be rare for the true
COSMOS ``twin'' of a Y1 source to have higher errors. 
The discarded pairs predominantly are cases of large COSMOS and small
Y1 galaxies (since large size raises flux errors), and the size
mismatch means these galaxies would  never be good matches.  Other discarded pairs come from
COSMOS galaxies lying in a shallow region of the DECam COSMOS
footprint, such as near a mask or a shallow part of the dither
pattern, and this geometric effect will not induce a redshift bias.}
Note
that the MOF fluxes used here make 
use of the SLR zeropoints, for both COSMOS and Y1 catalogs.  The size
metric is the one produced by \metacal.
\item The matched COSMOS2015 galaxy is selected as the one that
  minimizes the flux-and-size $\chi^2,$
\begin{equation}
\chi^2 \equiv \sum_{b\in griz} \left( \frac{f_b^{\rm Y1} - f_b^{\rm
      COSMOS}}{\sigma_b}\right)^2 + \left(\frac{s^{\rm Y1}-s^{\rm COSMOS}}{\sigma_s}\right)^2,
\end{equation}
where $f_b$ and $s$ are the fluxes in band $b$ and the size,
respectively, and $\sigma_b$ and $\sigma_s$ are the measurement errors
in these for the chosen source. We also find the galaxy that minimizes
the $\chi_{\rm flux}^2$ from flux differences only:
\begin{equation}
\chi_{\rm flux}^2 \equiv \sum_{b\in griz} \left( \frac{f_b^{\rm Y1} - f_b^{\rm
      COSMOS}}{\sigma_b}\right)^2\; .
\end{equation}
If the least $\chi_{\rm flux}^2$ is smaller than $(\chi^2-4)$ of the galaxy with the least flux-and-size $\chi^2$, we
use this former galaxy instead. Without this criterion, we could be using poor matches in flux (which is more predictive of redshift than size) by requiring a good size match (that does not affect redshift distributions much). 
It applies to about 15 per cent of cases.


\item A redshift $z^{\rm true}$ is assigned by drawing from the
$p^{\rm C30}(z)$ of the matched COSMOS2015 galaxy, using the rescaling of \S\ref{pdfrescale}.
\item A bin assignment is made by running the BPZ procedures of
  \S\ref{bpz} on the noise-matched $griz$ fluxes of the COSMOS
  match, using the mean value of each galaxy's posterior $p^{\rm PZ}(z)$, as before.  
  For the \imshape\ catalog, the MOF photometry of the COSMOS
  galaxy is used, just as is done for the Y1 main survey galaxies.  The \metacal\
  treatment is more complex: we generate simulated Metacal fluxes
  $f_b^{\rm meta, COSMOS}$ for  the COSMOS galaxy via
\begin{equation}
f_b^{\rm meta,COSMOS} = f_b^{\rm MOF,COSMOS} 
\frac{f_b^{\rm meta,Y1}}{f_b^{\rm MOF,Y1}}\;.
\end{equation}
\red{This has the effect of imposing on COSMOS magnitudes the same
difference between Metacal and MOF as is present in Y1, thus
imprinting onto COSMOS simulations any errors 
in the Y1 catalog \metacal\ magnitudes due to
neglect of the SLR or other photometric errors.}
For the flux uncertainty of these matched fluxes, for both MOF and Metacal, we assign the flux errors of the respective Y1 galaxy.

\item The effective weak lensing weight $w$ of the original source galaxy is
  assigned to its COSMOS match (cf.~\S\ref{binest}).
\end{enumerate}

\red{As a check on the matching process, we examined the distribution of
the $\chi_{\rm flux}^2$ between matched galaxies. The distribution is
skewed toward significantly lower $\chi^2$ values than expected from a 
true $\chi^2$ distribution with 4 degrees of freedom.  This indicates that
the COSMOS-Y1 matches are good: COSMOS galaxies are photometrically even more similar to the Y1 target galaxies than
they would be to re-observed versions of themselves.}

\begin{figure}
\centering
\includegraphics[width=\linewidth]{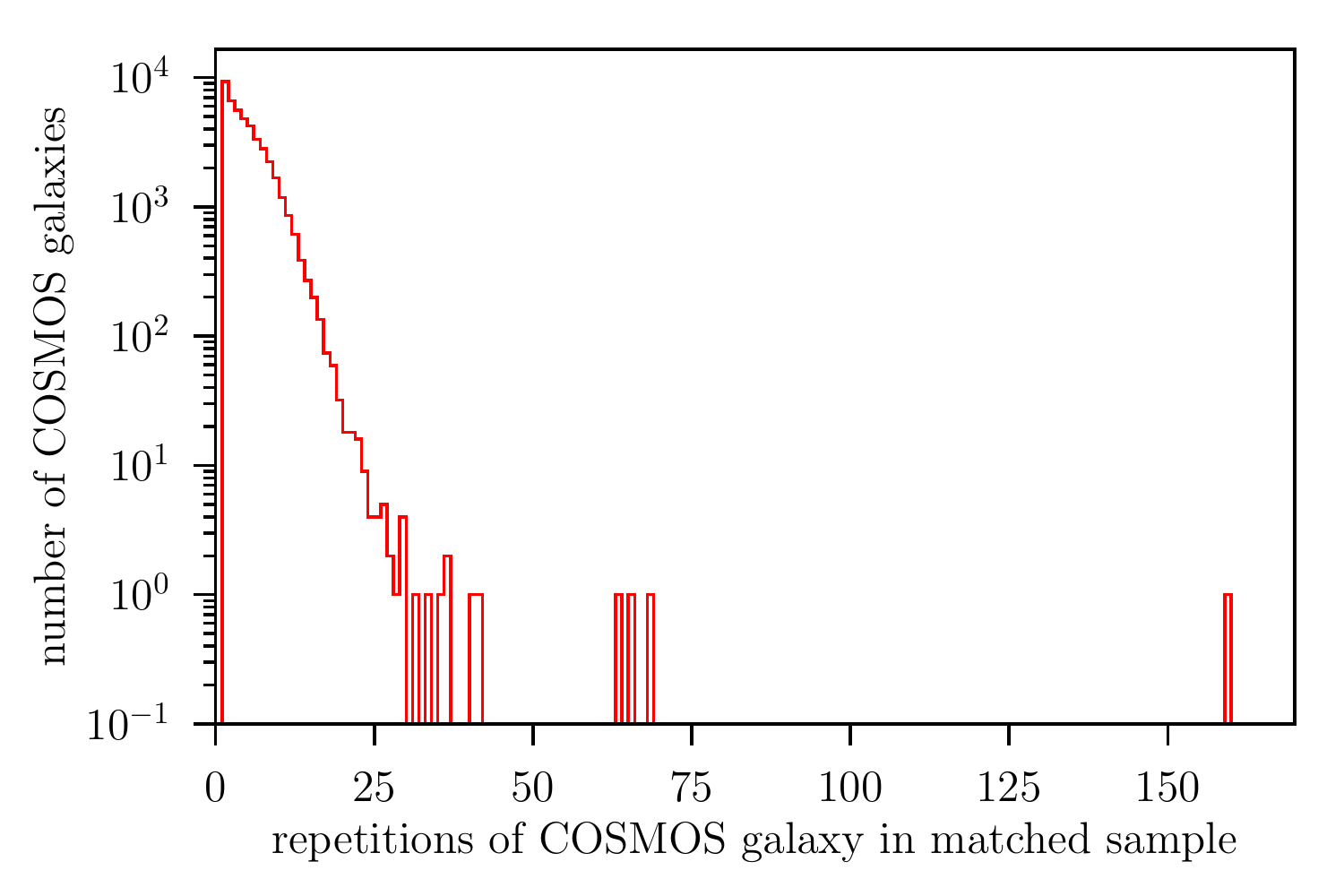}
\caption{Repetitions of COSMOS galaxies in the fiducial matched
  \metacal\ sample of 200,000 objects. The overall weight of galaxies
  with more than 20 matches, which are typically bright, is below one
  per cent of the total weight. The high-usage outliers are a few of the very bright
COSMOS galaxies.}
\label{fig:repetitions}
\gbR{Daniel, is this for a single resampling?}
\DGR{Yes, made clear}
\end{figure}

\red{A second check on the matching algorithm is to ask whether the
individual COSMOS galaxies are being resampled at the expected rates.
As expected, most sufficiently bright galaxies in COSMOS are used more
than once, while the faintest galaxies are used more rarely or
never. Figure \ref{fig:repetitions} shows the number of times each of the
COSMOS galaxies is matched to \metacal\ (if it is bright enough to be
matched at all) \red{in our fiducial matched catalog}.  We see that there is no unwanted tendency for a small fraction of the COSMOS galaxies to bear most of the resampling weight. All COSMOS galaxies with more than 50 repetitions are brighter than $i=18.5$ and have a typical redshift of $z\approx0.15$.}

We now use the matched COSMOS galaxy set to produce an estimate of the
difference in mean redshift between the $griz$-predicted distribution
and the ``truth'' provided by COSMOS2015 for all galaxies assigned
to a given source bin:
\begin{equation}
\Delta z = \frac{\sum_i w_i z^{\rm true}_i}{\sum_i  w_i} - \frac{\sum_j w_j z^{\rm PZ}_j}{\sum_j w_j} \; ,
\end{equation}
where the sums run over all matched COSMOS2015 galaxies $i$ and all galaxies in the original source sample $j$.

This construction properly averages $\Delta z$ over the observing conditions
(including photometric zeropoint errors) and weights of the Y1 WL
sources.   These estimated $\Delta z$ values \red{using BPZ} are tabulated in
Table~\ref{Dztable} for both WL source catalogs.

\begin{figure}
\centering
\includegraphics[width=\linewidth]{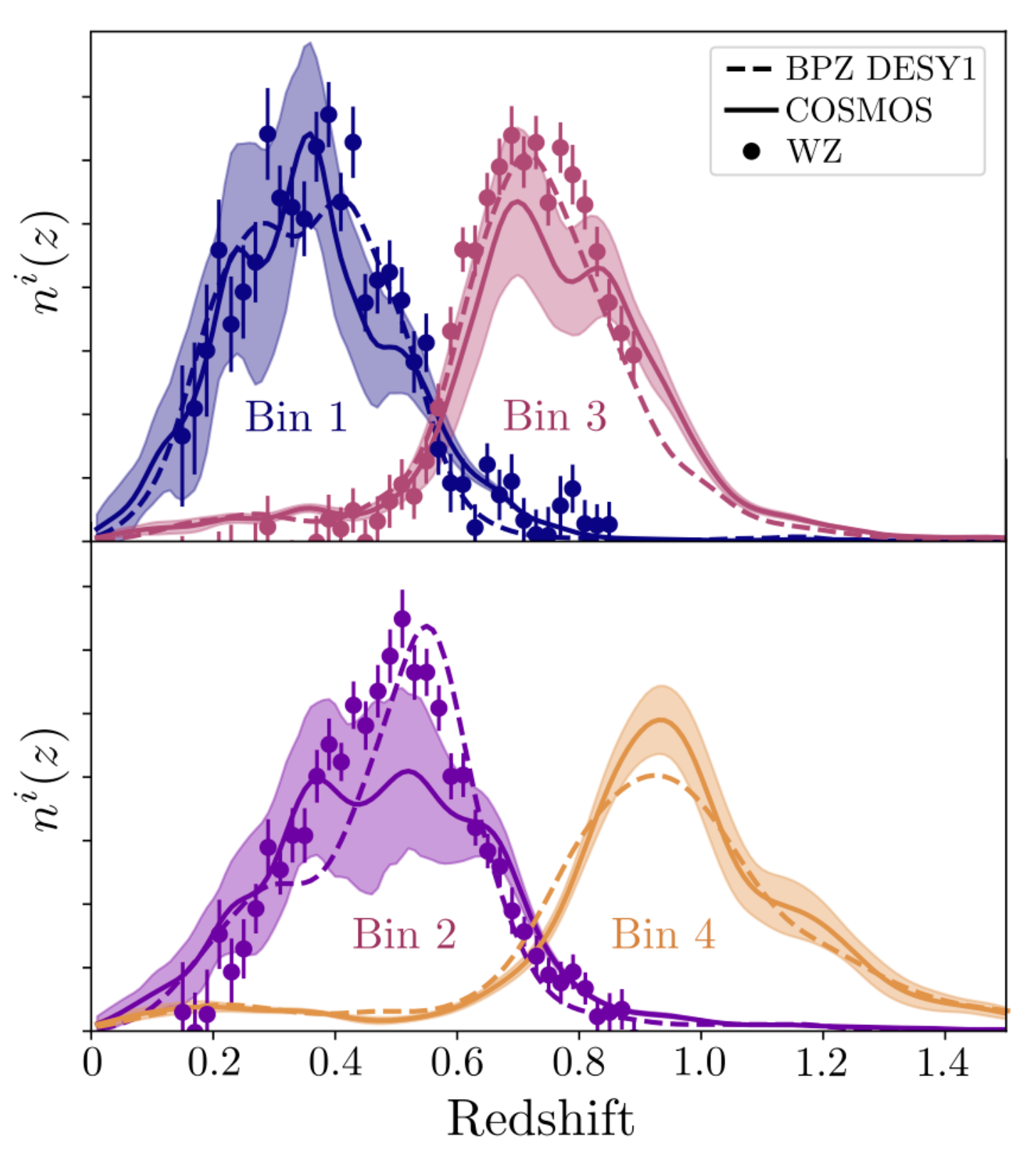}
\caption{The redshift distributions $n^i(z)$ derived from three
  different methods are plotted for each of the 4 WL \metacal\ source
  bin  populations $i=1\ldots4.$ The top (bottom) figure shows the 1st and 3rd (2nd and 4th) tomographic redshift bins. The clustering methodology (WZ) can
  only constrain $n^i(z)$ for $0.15<z<0.9$, and the normalization of
  the distribution is arbitrary for the bins extending beyond this
  range.  The band around the COSMOS $n^i(z)$ depicts the uncertainties as described in \S\ref{cosmos} and Table~\ref{Dztable}, and the error bars on WZ are statistical noise.
  There is some significant disagreement in the shapes of the
  distributions, particularly in $n^2(z).$  We demonstrate in \S\ref{dndzshapes} that
  this does not bias the DES Y1 cosmological inferences.}
\label{fig:dndz}
\end{figure}

The COSMOS validation also yields an estimate of $n^i(z)$ by a
weighted average of the rescaled $p^{\rm C30}(z)$'s of the matches (or,
equivalently, of samples drawn from them).
Figure~\ref{fig:dndz} plots these resampled-COSMOS estimates along with
the original $n^i_{\rm PZ}(z)$ from BPZ.  Here it is apparent that in
some bins, these two estimates differ by more than just a simple shift
in redshift---the shapes of the $n^i(z)$ distributions differ
significantly.  In \S\ref{dndzshapes} we demonstrate that these
differences do not bias our cosmological inferences.

\begin{table*}
\caption{Values of and error contributions to photo-$z$ shift
  parameters of BPZ $n^i(z).$}
\label{Dztable}
\begin{center}
\begin{tabular}{l|cccc|}
\hline
\hline
Value & Bin 1 & Bin 2 & Bin 3 & Bin 4  \\
\hline 
$z^{\rm PZ}$ range & 0.20--0.43 & 0.43--0.63 & 0.63--0.90 & 0.90-1.30\\[3pt]

COSMOS footprint sampling & 
$\pm0.0073$ &
$\pm0.0077$ &
$\pm0.0039$ &
$\pm0.0070$ \\

COSMOS limited sample size & 
$\pm0.0009$ &
$\pm0.0017$ &
$\pm0.0018$ &
$\pm0.0030$ \\

COSMOS photometric calibration errors & 
$\pm0.0030$ &
$\pm0.0040$ &
$\pm0.0039$ &
$\pm0.0059$ \\

COSMOS hidden variables &
$\pm0.0066$ &
$\pm0.0066$ &
$\pm0.0066$ &
$\pm0.0066$ \\

COSMOS errors in matching &
$\pm0.0073$ &
$\pm0.0073$ &
$\pm0.0073$ &
$\pm0.0073$ \\ [5pt]

COSMOS \red{single-bin} $\Delta z^i$ uncertainty & 
$\pm0.013$ & $\pm0.013$ & $\pm0.011$ & $\pm0.014$ \\[5pt]

\multicolumn{5}{c}{\metacal} \\[3pt]
 
COSMOS final $\Delta z^i$, \red{tomographic uncertainty} & 
$-0.006\pm0.020$ & 
$-0.014\pm0.021$ & 
$+0.018\pm0.018$ & 
$-0.018\pm0.022$ \\

WZ final $\Delta z^i$ & 
$+0.007\pm0.026$ & 
$-0.023\pm0.017$ & 
$+0.003\pm0.014$  & 
--- \\
\textbf{Combined final $\Delta z^i$} & 
$-0.001 \pm0.016$ &
$-0.019 \pm0.013$ &
$+0.009 \pm0.011$ &
$-0.018 \pm0.022$ \\[5pt]

\multicolumn{5}{c}{\imshape} \\[3pt]
 
COSMOS final $\Delta z^i$, \red{tomographic uncertainty} & 
$+0.001\pm0.020$ & 
$-0.014\pm0.021$ & 
$+0.008\pm0.018$ & 
$-0.057\pm0.022$ \\

WZ final $\Delta z^i$ & 
$+0.008\pm0.026$ & 
$-0.031\pm0.017$ & 
$-0.010\pm0.014$ & 
--- \\ 

\textbf{Combined final $\Delta z^i$} & 
$+0.004 \pm0.016$ &
$-0.024 \pm0.013$ &
$-0.003 \pm0.011$ &
$-0.057 \pm0.022$ \\
\hline
\end{tabular}
\end{center}
\end{table*}

In the following subsections, we determine several contributions to
the uncertainty of these $\Delta z^i$. All of these are presented for
the \metacal\ sample binned by BPZ redshift estimates. For \imshape\
galaxies with BPZ, we use the same uncertainties. Results for DNF are
in Appendix~\ref{dnfresults}.

From the resampling procedure, we also determine common metrics on the photo-$z$ performance in \autoref{standard_metrics}.

\subsection{Sample variance contribution}
\label{sec:err:cv}
The first contribution to the uncertainty in the COSMOS $\Delta z^i$'s
is from sample variance from the small angular size of the COSMOS2015
catalog.  Any attempt at analytic estimation of this uncertainty would
be complicated by the reweighting/sampling procedure that alters the
native $n(z)$ of the COSMOS line of sight, so we instead estimate the
covariance matrix of the $\Delta z^i$ by repeating our procedures on
different realizations of the COSMOS field in the Buzzard simulated
galaxy catalogs.
 
\begin{figure}
\centering
\includegraphics[width=\linewidth]{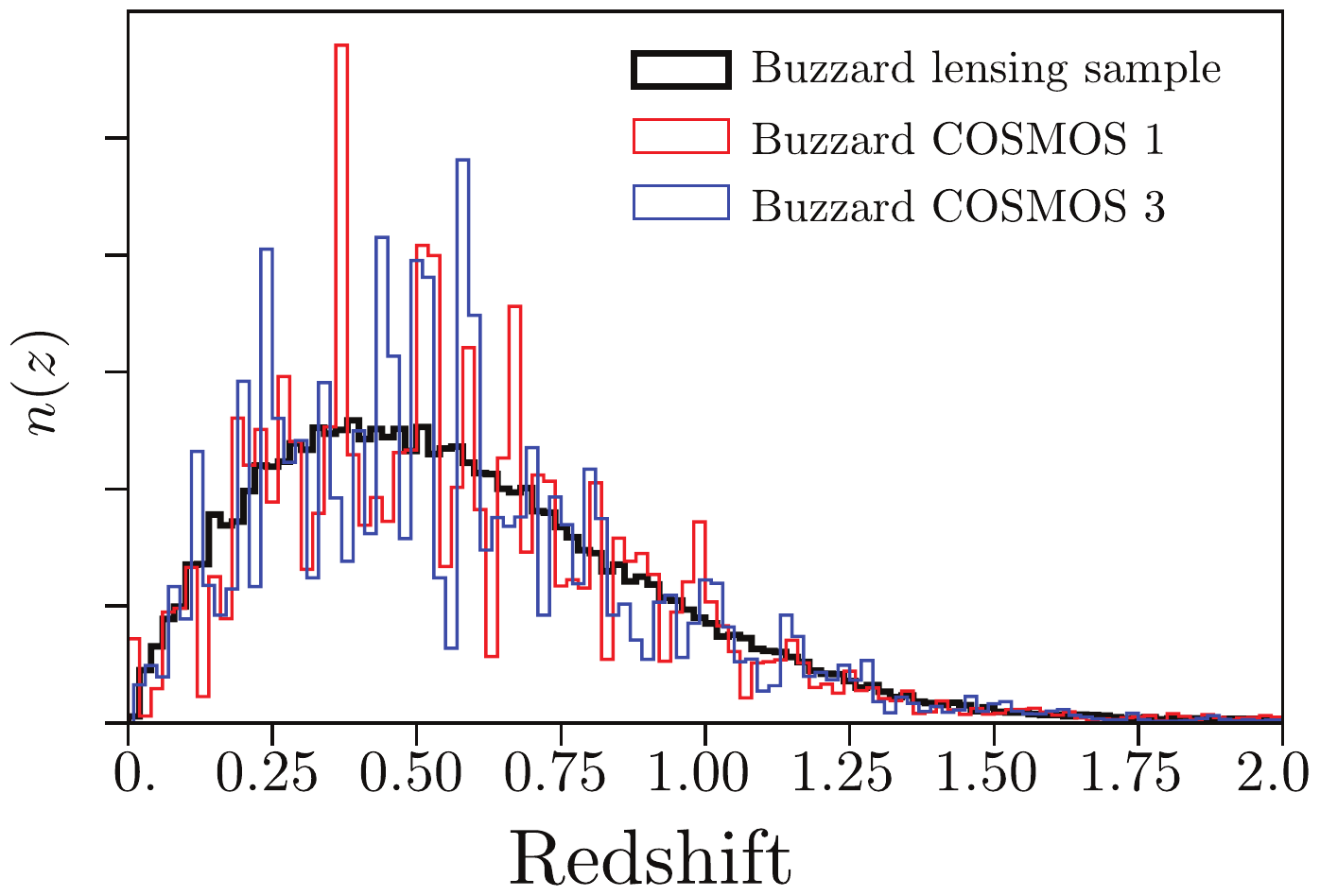}
\caption{Redshift distributions of the full simulated lensing sample
  from the Buzzard catalog
  (grey) and two examples of samples from COSMOS-sized footprints in
  the Buzzard catalogs that have been resampled and weighted to
  match the full distribution (blue and orange).}
\label{fig:buzzardresampling}
\gbR{We'll keep this figure, but it needs to be readable.  Convert
  the histograms from filled to different-colored outlines so they can
  be seen despite overlap.}
  \DGR{done}
\end{figure}

\begin{figure}
\centering
\includegraphics[width=0.8\linewidth]{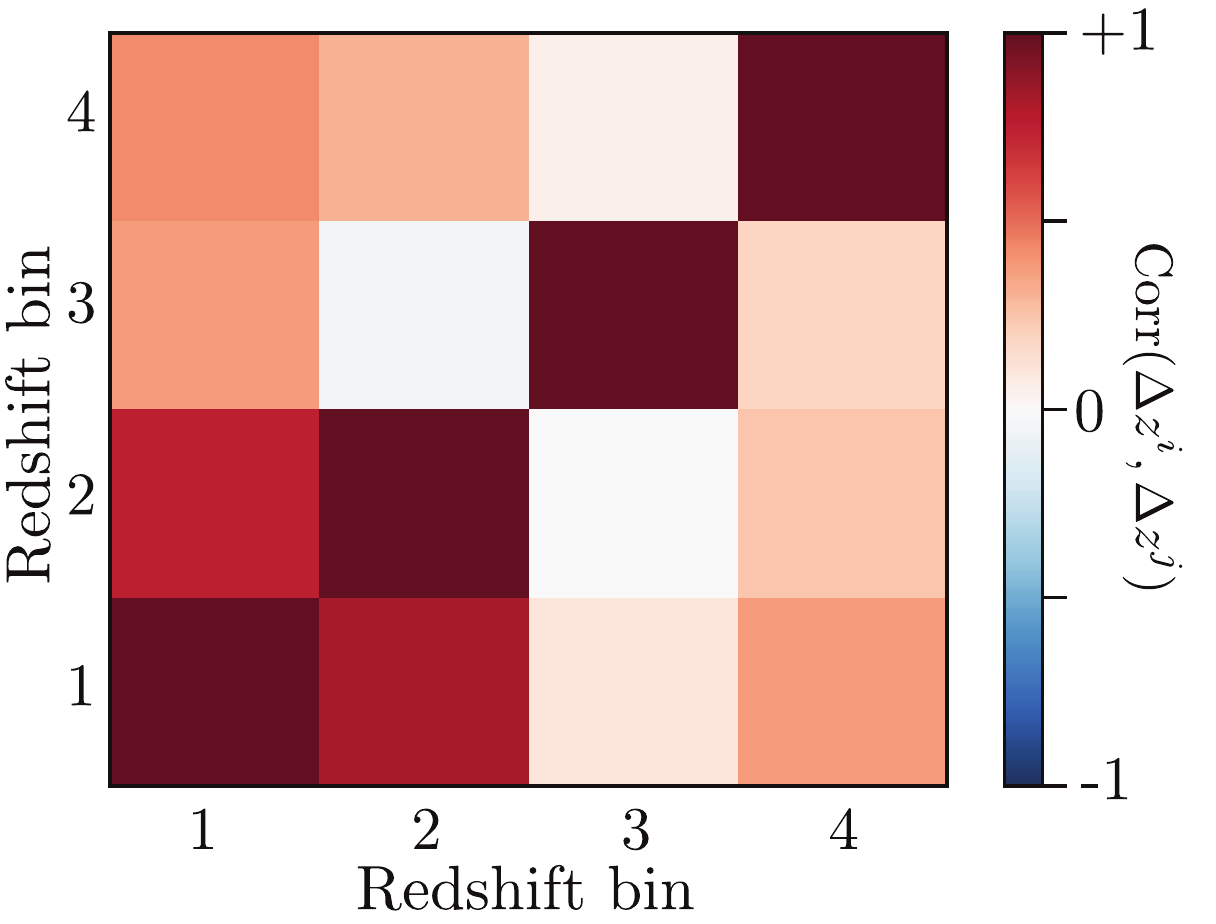}
\caption{Correlation coefficients of error on $\Delta z^i$ due to sample variance in COSMOS-resamplings between our four source redshift bins. Shown is the correlation matrix for the \metacal\ sample binned by BPZ.}
\label{fig:buzzardcorrelation}
\end{figure}

The resampling procedure of \S\ref{cosmosresample} is repeated
using a fixed single draw of 200,000 galaxies from a Buzzard
simulated Y1 WL sample (\S\ref{buzzard}) as catalog A, and
367 randomly placed COSMOS-shaped cutouts from the Buzzard truth catalog, i.e.~a catalog with noiseless flux information, as
catalog B. Each of these yields an independent $n(z)$ of the matched COSMOS catalogs (cf.~\autoref{fig:buzzardresampling}), and consequently an independent sample variance realization of the $\Delta z^i$. 
There are significant correlations between the $\Delta z^i$ bins,
especially bins 1 and 2, as shown in \S\ref{fig:buzzardcorrelation}.
The diagonal elements are listed 
as ``COSMOS footprint sampling'' in Table~\ref{Dztable}.




Since we use the same subset of the Buzzard lensing sample for
  each of the COSMOS-like resamplings, this variance estimate
  does not include the uncertainty due to the limited subsample size
  of 200,000 galaxies. We estimate the latter effect by \red{resampling of
  the $\Delta z^i$ in this sample,} and find it to be
  subdominant ($\sigma_{\Delta z}^i <0.003$ in all redshift bins,
  ``limited sample size'' in Table~\ref{Dztable}).


\subsection{Photometric calibration uncertainty}
\label{sec:err:slr}
The $griz$ DECam photometry of the COSMOS field has 
uncertainties in its zeropoint due to errors in the
SLR-based calibration. While the Y1 catalog averages over the SLR
errors of many fields, the validation is sensitive to the single
realization of SLR errors in the COSMOS field.  We estimate the
distribution of zeropoint errors by comparing the SLR zeropoints in
the Y1 catalog to those derived from the superior ``forward global
calibration module'' (FGCM) \red{and reddening correction} applied to three years' worth of DES
exposures by \citet{fgcm}. In this we only use regions with Galactic extinction $E(B-V)<0.1$, since the COSMOS field has relatively low extinction and strong reddening might cause larger differences between the FGCM and SLR calibration.
The root-mean-square zeropoint offsets between SLR and FGCM calibration are between 0.007 ($z$) and 0.017 ($g$). 

We estimate the
impact on $\Delta z^i$ by drawing 200 mean-subtracted samples of photometric offsets
from the observed (FGCM-SLR) distribution, applying each to the COSMOS
fluxes, and repeating the derivation of $\Delta z^i$ as per
\S\ref{cosmosresample}.  
Table \ref{Dztable} lists
the uncertainty of the $\Delta z^i$ of each of the four tomographic
bins due to those, which are $0.003-0.006$.

\gbR{removing the photo-calibration error figure.}

\subsection{Hidden-variable uncertainty}
\label{sec:err:hidden}
We have matched COSMOS galaxies to the shear catalog galaxies by their
$griz$ fluxes and by their estimated pre-seeing size. This set of
parameters is likely not completely predictive of a galaxy's
selection and weight in our shear catalog.
Other morphological properties (such as the steepness of its profile)
probably matter and do correlate with redshift \citep[e.g.][]{Soo2017}. In addition, the matching in size is only done in 85 per cent of cases to begin with \autoref{cosmosresample}. 

To estimate the effect of any variables hidden to our matching
algorithm, we repeat the process while ignoring the size variable.
We find changes in $\Delta z^i$ for \metacal\ to be
$(+0.010,+0.015,+0.009,+0.014)$ in the four bins. \cite{Soo2017} found that the single morphological parameter that provides the greatest improvement in $\sigma_{68}$ and outlier fraction is galaxy size. Since we therefore expect the size to have the strongest influence on both lensing and redshift, and we
\emph{are} correcting for size, we estimate the potential influence of
any further variables as no more than half of the size effect. 
\RR{We do not assume that these systematic errors found in
simulations are exactly equal in the data - rather, we only assume
that the two are of similar size, and thus use the rms of offsets
found in the simulation as the width of a Gaussian systematic
uncertainty on the data.}
We take half of the quadratic mean of the shifts in the four redshift bins, $\pm0.0066,$ as our estimate of 
the hidden-variable uncertainty in each bin. 
These biases are
  likely to be correlated between bins.  In \S\ref{sec:bincorr} we describe a
  modification to our single-bin uncertainties that accounts for potential correlations.

\subsection{Systematic errors in matching}

Even in the absence of the above uncertainties, the resampling
algorithm described above might not quite reproduce the true redshift
distribution of the input sample. 
The matching algorithm may not, for example, pick a COSMOS galaxy
  which is an unbiased estimator of the target galaxy's redshift,
  especially given the sparsity and inhomogeneous distribution of the COSMOS
  sample in the four- to five-dimensional space of $griz$ fluxes and
  size.

We estimate the size of this effect on $\Delta z^i$ using the \emph{mean} offset in binned mean true redshift of the 367 realizations of resampled COSMOS-like catalogs in the Buzzard simulations (see~\autoref{sec:err:cv}) from the binned mean true redshift of the underlying Buzzard shape sample. 

We find differences in mean true redshift of sample A -
  matched B of (0.0027, 0.0101, 0.0094, 0.004) for the four redshift bins. Since the simulation
  is not fully realistic, we do not attempt to 
  correct the result of our resampling with these values. Rather, we take
  them as indicators of possible systematic
  uncertainties of the resampling algorithm. \RR{Following the argument in section \S \ref{sec:err:hidden} ,we} thus use the
  quadratic mean of these values (0.0073) as a systematic
  uncertainty in each bin.


\subsection{Combined uncertainties and correlation between redshift bins}
\label{sec:bincorr}

The final uncertainties on the $\Delta z^i$ are estimated by adding
in quadrature the contributions listed above, yielding the ``COSMOS total $\Delta z^i$ uncertainty'' 
in Table~\ref{Dztable}.
These values are derived independently for each redshift bin, but it
is certain that the $\Delta z^i$ have correlated errors, e.g. from
sample variance as shown in Figure~\ref{fig:buzzardcorrelation}, and
such correlations should certainly be included in the inference of
cosmological parameters.
The values of the off-diagonal elements of the combined COSMOS $\Delta z$
covariance matrix, are, however, difficult to estimate with any precision.  In
Appendix~\ref{correlationappendix} we demonstrate that by increasing
the diagonal elements of the covariance matrix by a factor $(1.6)^2$
and nulling the off-diagonal elements, we can ensure that any
inferences based on the $\Delta z^i$ are conservatively estimated
for any reasonable values of the off-diagonal elements.  We therefore
apply a factor of 1.6 to all of the single-bin uncertainties in
deriving the ``COSMOS final $\Delta z^i$'' constraints for \metacal\
and \imshape\ given in Table~\ref{Dztable}.

\subsection{Standard photo-z performance metrics}
\label{standard_metrics}

\begin{table*}
\caption{Common performance metrics and uncertainties measured using BPZ point predictions and draws from the rescaled COSMOS2015 PDFs. \RR{The quantity $\sigma_{68}(R)$ is the 68\% spread of the residual distribution $R$, about the median. }The outlier fraction is defined as the fraction of galaxies with $griz$ redshift estimates than $2\times\sigma_{68}(R)$ from the COSMOS2015 value.}
\label{bpz_table_metrics}
\begin{center}
\begin{tabular}{l|cccc|}
\hline
metric & $0.20<z<0.43$ & $0.43<z<0.63$ & $0.63<z<0.90$ & $0.90<z<1.30$\\ \hline
\multicolumn{5}{c}{BPZ \metacal\ binning, MOF $p^{\rm PZ}(z)$} \\[3pt]
$\sigma_{68}\RR{(R)}$ & $0.12 \pm 0.01$ & $0.16 \pm 0.01$ &$0.12 \pm 0.01$ &$0.17 \pm 0.01$  \\
Outlier Fraction \% & $3.3 \pm 0.5$ &$3.6 \pm 0.8$ &$6.1 \pm 0.4$ &$6.6 \pm 0.5$  \\  \hline
\multicolumn{5}{c}{DNF \metacal} \\[3pt]
$\sigma_{68}\RR{(R)}$ & $0.10 \pm 0.01$ &$0.16 \pm 0.01$ &$0.16 \pm 0.01$ &$0.21 \pm 0.01$  \\
Outlier Fraction \% & $5.0 \pm 0.4$ &$3.8 \pm 0.8$ &$8.6 \pm 0.6$ &$7.5 \pm 0.5$\\
\end{tabular}
\end{center}
\end{table*}

Although not a critical input to the cosmological tests of \citet{keypaper},
we determine here some standard metrics of photo-$z$ performance.
We define the residual
$R$ as the difference between the mean of the $p^{\rm PZ}(z)$
using the MOF photometry and a random draw from the COSMOS
$p^{\rm C30}(z)$ matched during resampling. We use a random draw from
$p^{\rm C30}(z)$ rather than the peak, so that uncertainty in these
``truth'' $z$'s is included in the metrics. \magenta{Because the width of $p^{\rm C30}(z)$ is much smaller than that of $p^{\rm PZ}(z)$, this does not affect the results significantly.}
\WGHR{Perhas we should mention the avergae 68\%ile width of COSMOS PDF, so that the reader can mentally subtract it from our metric values?}
\DGR{Added statement on this. I think Ben did remove the C30 width in quadrature at the end, but it doesn't change the numbers significantly.}

We define $\sigma_{68}(R)$ as the 68\% spread of $R$ around its
median. In this section $\sigma_{68}(R)$ measures the departure of the
mean of $p^{\rm PZ}(z)$ from the true $z$, whereas the
$\sigma_{68}$ in Figure~\ref{sigma_bpz} is a measure of width of
$p^{\rm PZ}(z)$ independent of any truth redshifts. We also measure
the outlier fraction, defined 
as the fraction of data for which $|R| > 2\times\sigma_{68}$. If the
redshift distribution were Gaussian, the outlier fraction would be
5\%, and this metric is a measure of the tails of the $R$
distribution. 

We calculate the uncertainties on these metrics \red{from sample
  variance, COSMOS photometric calibration uncertainty, and selection
  of the lensing sample by hidden variables
  (cf. \autoref{sec:err:cv}-\ref{sec:err:hidden}).}
We add each of the these uncertainties in quadrature in each tomographic bin, and highlight that the largest source of uncertainty is due to sample variance.

Table~\ref{bpz_table_metrics} presents the metric values and
uncertainties of the galaxies in each redshift bin, using the
\metacal\ sample and binning.

\section{Combined constraints}
\label{res}
To supplement the constraints on $\Delta z^i$ derived above using the
COSMOS2015 photo-$z$'s, we turn to the ``correlation redshift''
methodology \citep{newman,menard,schmidt} whereby one measures the
angular correlations between the unknown sample (the WL sources) and
a population of objects with relatively well-determined redshifts.  In
our case the known population are the \redmagic galaxies, selected
precisely so that their $griz$ colors yield high-accuracy photometric redshift
estimates. 

An important complication of applying WZ to DES Y1 is that we do not have a sufficient sample of galaxies with known redshift available that spans the redshift range of the DES Y1 lensing source galaxies -- the \redmagic\  galaxies do not extend beyond $0.2<z<0.9$. Constraints on the mean redshift of a source population can still be derived in this case, but only by \emph{assuming} a shape for the $n(z)$ distribution, whose mean is then determined by the clustering signal in a limited redshift interval. A mismatch in shape between the assumed and true $n(z)$ is a source of systematic uncertainty in such a WZ analysis. One of the main results of \citealt{xcorrtechnique}, which describes the implementation and full estimation of uncertainties of the WZ method for DES Y1 source
galaxies, is that while this systematic uncertainty needs to be accounted for, it is not prohibitively large. This statement is validated in \citealt{xcorrtechnique} for the degree of mismatch between the true $n(z)$ and the $n(z)$ found in a number of photometric redshift methods applied to simulated galaxy catalogs. The redshift distributions of the DES weak lensing sources as estimated by BPZ, as far as we can judge this from the comparison with the COSMOS estimates of their true $n(z)$, show a similar level of mismatch to the truth. The systematic uncertainty budget derived in \citet{xcorrtechnique} is therefore applicable to the data. We do not, however, attempt to correct the systematic offsets in WZ estimates of $\Delta z^i$ introduced due to this effect -- for this, we would require the galaxy populations and photometric measurements in the simulations to be perfectly realistic.

\begin{figure*}
\centering
\includegraphics[width=0.9\textwidth]{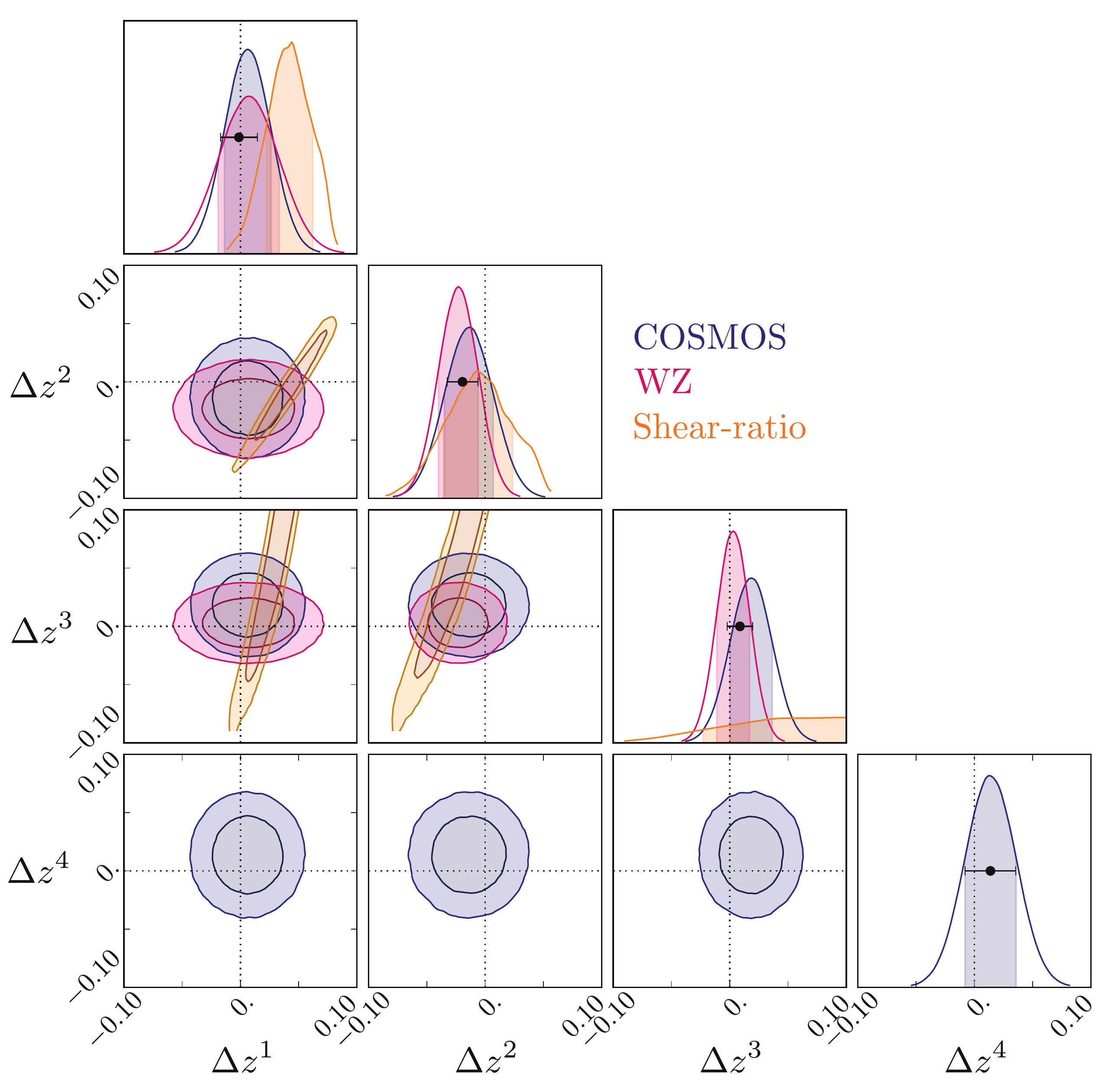}
\caption{Constraints on the shifts $\Delta z^i$ applied to
  the \metacal\ $n_{\rm PZ}(z)$ distributions for the weak lensing source
  galaxies are plotted for three different 
  validation techniques.  Shifts derived from resampling the COSMOS
  30-band redshifts are described in this paper, and agree well with
  those derived (for bins 1--3 only) using angular correlations
  between the source population and redMagic galaxies (WZ) by
  \citet{xcorr} (COSMOS constraints plotted here have been expanded as
  per Appendix~\ref{correlationappendix} to include the effects of
  poorly known correlation between bins).
These are also consistent with the weak lensing shear
  ratio tests conducted by \citet{gglpaper}.  The
  final validation constraints 
  on $\Delta z^i$ are taken as the combination of the COSMOS and WZ
  results for each redshift bin (where available), and yield the 68\%
  confidence intervals denoted by the \magenta{black points and error bars} in the 1-d marginal
  plots.  The dashed lines at $\Delta z^i=0$ indicate no mean shift
  from the BPZ posteriors---the validation processes yield shifts 
  that are non-zero at $\approx 1\sigma$ level.
}
\label{deltaz2d}
\end{figure*}

The method is applied to DES Y1 data in \citet{xcorr}. A similar analysis was performed on the DES SV data set in \citet{2017arXiv170708256D}.
The
resultant estimates of $\Delta z^i$ are listed in Table~\ref{Dztable}
and plotted in Figure~\ref{deltaz2d}.  
The full $n^i(z)$'s estimated from the WZ
method are plotted in Figure~\ref{fig:dndz}. Note that the WZ method obtains
no useful constraint for bin 4 because the \redmagic sample is
confined to $z<0.9$ and thus has little overlap with bin 4. Due to the lack of independent confirmation, the redshift calibration of this bin should be used with greater caution -- in \citet{keypaper} and \citet{2017arXiv171005045G}, we indeed show that constraints do not significantly shift when the bin is removed from the analysis. 

In the three lower redshift bins, the COSMOS and WZ validation methods generate estimates of $\Delta
z^i$ that are fully consistent.  Indeed even their
$n^i(z)$ curves show qualitative agreement.  We therefore proceed to
combine their constraints on $\Delta z^i$ to yield our most accurate
and reliable estimates.  The statistical errors of the COSMOS and WZ
methods are uncorrelated (sample variance in the COSMOS
field vs. shot noise in the measurements of angular correlations in the
wide field).  The dominant systematic errors of the two methods should
also be uncorrelated, e.g. shortcomings in our resampling for COSMOS vs.
uncertainties in the bias evolution of source galaxies for WZ. We are
therefore confident that we can treat the COSMOS and WZ constraints as
independent, and we proceed to combine them by multiplying their
respective 1-dimensional Gaussian distributions for each $\Delta z^i,$
i.e. inverse-variance weighting.  In bin 4, the final constraints are
simply the COSMOS constraints since WZ offers no information.

The resultant constraints, listed for both \metacal\ and \imshape\
catalogs in Table~\ref{Dztable}, are the principal result of this
work, and are adopted as input to the cosmological inferences of
\citet{shearcorr} and \citet{keypaper}.  
The adopted 68\%-confidence ranges
for each $\Delta z^i$ are denoted by the gray bands in the 1-d
marginal plots of Figure~\ref{deltaz2d}.  

\red{One relevant question is whether our calibration finds that significant non-zero shifts are required to correct the photo-$z$ estimates of the mean redshift. For the fiducial \metacal\ BPZ, this is not the case: the $\chi^2=\sum_i(\Delta z^i/\sigma_{\Delta z^i})^2$ is 3.5 with 4 bins. However, the combined $\Delta z^4$ is non-zero at $2.6\sigma$ for \imshape\ BPZ and the $\Delta z^2$ is non-zero at $3.3\sigma$ for \metacal\ DNF,
indicating that there are significant alterations being made to \red{some of} the
$n_{\rm PZ}^i(z)$ estimates.}

A further check of the accuracy of our $n^i(z)$ estimation is
presented by \citet{gglpaper} using the ratios of lensing shear on the
different source bins induced by a common set of lens galaxies.
Initially proposed as a cosmological test \citep{jaintaylor}, the
shear ratio is in fact much less sensitive to cosmological parameters
than to potential errors in either the calibration of the shear
measurement or the determination of the $n^i(z).$  We plot
in Figure~\ref{deltaz2d} the constraints on $\Delta z^i$ inferred by
\citet{gglpaper} after marginalization over the estimated errors in shear
calibration and assuming a fixed $\Lambda$CDM cosmology with $\Omega_{\rm m}=0.3$.  The
shear-ratio test is fully consistent with the COSMOS and WZ estimates
of $\Delta z^i,$ though we should keep in mind that this test is also
dependent on the validity of the shear calibration and some other
assumptions in the analysis, \RR{and importantly is covariant with the WZ method, because both methods
rely on correlation functions as measured with respect to the same galaxy samples.}.

\section{Use for cosmological inference}
\label{inference}
The final rows of Table~\ref{Dztable} provide the prior on errors in
the redshift distributions used during inference of cosmological
parameters for the DES Y1 data, under the assumption that errors in
the $n^i(z)$ resulting from the photo-$z$ analysis follow \eqq{Dz}.  
Determination of redshift distributions is and will continue to be one
of the most difficult tasks for obtaining precision cosmology from
broadband imaging surveys such as DES, so it is important 
to examine the potential
impact of assumptions in our analysis choices. Further, we wish to
identify areas where our methodology can be improved and thereby
increase the precision and accuracy of future cosmological analyses. 

\subsection{Dependence on COSMOS2015 redshifts}
\label{laigleerrors}
\label{sec:err:spec}

First, we base our COSMOS validation on the COSMOS2015 redshift
catalog derived from fitting spectral templates to 30-band fluxes.
Our COSMOS validation rests on the assumption that \citet{Laigle} have
correctly estimated the redshift posteriors of their sources.  
Overall, redshift biases in the COSMOS2015 redshifts are
significant, unrecognized sources of error in our cosmological
inferences if they approach or
exceed the $\delta z\approx0.01$--0.02 range of uncertainty in our
$\Delta z^i$ constraints.  More precisely, this bias must accrue to the
portion of the COSMOS2015 catalog that is bright enough to enter
the DES Y1 shear catalogs.  

For the subset of their sources with spectroscopic redshifts, 
\citet{Laigle} report that galaxies in
the magnitude interval $22<i<23$ have ``catastrophic'' disagreement
between photo-$z$ and spectroscopic $z$ for only 1.7\% (0.6\%) for
star-forming (quiescent) galaxies (their Table~4).  This is the
magnitude range holding the 50\% completeness threshold of the DES Y1
shear catalogs. Brighter bins have
lower catastrophic-error rates, and only about 5 per cent of weight in the \metacal\ lensing catalog is provided by galaxies fainter than $i=23$. It would thus be difficult for these catastrophic
errors to induce photo-$z$ errors of 0.01 or more.  

\magenta{About 30 per cent of the galaxies used for the COSMOS weak lensing validation have spectroscopic redshifts from the latest 20,000 $I<22.5$ selected zCOSMOS DATA Release DR3, covering $1.7$~deg$^2$ of the COSMOS field to $z<1.2$ \citep{2009ApJS..184..218L}. We can thus use this subsample as an additional test of this statement. In all redshift bins, the shifts in the mean redshift estimated using this spectroscopic subset are very similar (less than 1-sigma of our error estimate) to the corresponding shifts estimated with photometric redshifts in the full sample. The difference between the 30-band (corrected) photometric mean redshifts and the corresponding spectroscopic redshifts for this subset is also within our error estimates.  
These tests indicate that the potential (unknown) biases in the 30-band photometric redshifts are smaller than other sources of uncertainty in the mean redshifts used for our WL analysis.}

Of greater concern
is the potential for bias in the portion of the DES detection regime
for which spectroscopic validation of COSMOS2015 photo-$z$'s is not
possible.  Neither we nor \citet{Laigle} have direct validation of this
subsample, so we are relying on the success of their template-based
method and broad spectral coverage in the spectroscopic regime to
extend into the non-spectroscopic regime.  Our confidence is boosted,
however, by the agreement in $\Delta z^i$ between the COSMOS
validation and the independent WZ validation in bins 1, 2, and 3.

\RR{Finally we note that we have also attempted to validate the photo-z distributions using only the galaxies with spectroscopic redshifts in the COSOMOS field, and find consistent, albeit uncompetitive results. The number of galaxies with spectra (20k) is an order of magnitude less than those with reliable photometric redshifts which increases statistical uncertainties, cosmic variance uncertainties and uncertainties from data re-weighting.}

\subsection{Insensitivity to $n^i(z)$ shape}
\label{dndzshapes}
\eqq{Dz} assumes that the only errors in the $n_{\rm PZ}^i(z)$
distributions take the form of a translation of the distribution in
redshift.  We do not expect that errors in the photo-$z$ distribution
actually take this form; rather we assume that the shape of $n^i(z)$
has little impact on our cosmological inference as long as the mean of
the distribution is conserved---and our methodology forces the mean of
the $n^i(z)$ to match that derived from the COSMOS2015 resampling.
The validity of this assumption can be tested by 
assuming that any errors in the shifted-BPZ $n^i(z)$ from \eqq{Dz} are
akin to the difference between these distributions and $n^i_{\rm
  COSMOS}(z)$ derived from the resampled COSMOS catalogs during the
validation process of \S\ref{cosmosresample}.  We produce a simulated
data vector for the DES Y1 cosmology analysis of \citet{keypaper} from a
noiseless theoretical prediction using the $n^i_{\rm COSMOS}(z)$
distributions.  We then fit this data using a model that assumes the
shifted BPZ distributions.  The best-fit cosmological parameters
depart from those in the input simulation by less than ten per cent of the uncertainty of \citet{keypaper}. We therefore confirm
that the detailed shape of $n^i(z)$ is not important to the Y1
analysis.

\subsection{Depth variation}
A third assumption in our analysis is that the $n^i(z)$ are the same
for all portions of the DES Y1 catalog footprint (aside, of course,
from the intrinsic density fluctuations that we wish to measure).
This is not the case: the failure to apply SLR \RR{adjustment} to our
fluxes in BPZ (\S\ref{knownerrors}) means that we have not corrected
our Metacal fluxes for Galactic extinction, and therefore have angular
variations in survey depth and photo-$z$ assignments.  Even without
this error, we would have significant depth fluctuations because of
variation in the number and quality of exposures on different parts of
the survey footprint.  

Appendix~\ref{inhomogeneity} provides an approximate quantification of the
impact of $n^i(z)$ inhomogeneities on our measurements of the 2-point
correlation functions involving the shear catalog.  There we conclude
that the few per cent fluctuations in survey depth and color calibration that
exist in our source catalogs should not significantly influence our
cosmological inferences, as long as we use the
source-weighted mean $n^i(z)$ over the survey footprint.  Both the
COSMOS and WZ validation techniques produce source-weighted
estimates of $\Delta z^i,$ as required.
\WGHR{para sounded like the appendix calc. didn't include effect of SLR issue on color. Modified slightly.}

\section{Conclusions and future prospects}
\label{conclusions}
\magenta{We have estimated redshift distributions and defined
  tomographic bins of source galaxies in DES Y1 lensing analyses from
  photometric redshifts based on their $griz$ photometry. While we use
  traditional photo-$z$ methods in these steps, we independently
  determine posterior probability distributions for the mean redshift
  of each tomographic bin that are then used as priors for subsequent lensing analyses.} 

The method for determining these priors developed in this paper is to match galaxies with COSMOS2015 30-band photometric redshift estimates to DES Y1 lensing source galaxies, selecting and weighting the former to resemble the latter in their $griz$ flux and pre-seeing size measurements. The mean COSMOS2015 photo-$z$ of the former sample is our estimate of the mean redshift of the latter. We determine uncertainties in this estimate, which we find have comparable, dominant contributions from 
\begin{enumerate}
\item sample variance in COSMOS, i.e.~the scatter in measured mean redshift calibration due to the limited footprint of the COSMOS field,
\item the influence of morphological parameters such as galaxy size on
  the lensing source sample selection, and
\item systematic mismatches of the original and matched sample in the algorithm we use.
\end{enumerate}
A significant reduction of the overall uncertainty of the mean redshift priors derived in this work would thus only be possible with considerable additional observations and algorithmic advances.

Subdominant contributions, in descending order, are due to
\begin{enumerate}
\item errors in photometric calibration of the $griz$ data in the COSMOS field and
\item the finite subsample size from the DES Y1 shear catalogs that we use for resampling. 
\end{enumerate}

The COSMOS2015 30-band photometric estimates of the mean redshifts, supplemented by consistent
measures by means of angular correlation against DES \redmagic galaxies in all but the highest redshift bins \citep{xcorrtechnique,redmagicpz,xcorr}, have
allowed us to determine the mean redshifts of 4 bins of WL source
galaxies to 68\% CL accuracy $\pm0.011$--0.022, independent of the
original BPZ redshift estimates used to define the bins and the
nominal $n_{\rm PZ}^i(z)$ distributions.  These
redshift uncertainties are a highly subdominant contributor
to the error budget of the DES Y1 cosmological parameter
determinations of \citet{keypaper} when marginalizing over the full set of nuisance parameters.  Likewise, the methodology of marginalizing over the mean redshift uncertainty only, rather than over the full shape of the $n^i(z)$, biases our analyses at less than ten per cent of their uncertainty.
Thus the methods
and approximations herein are sufficient for the Y1 analyses.

DES is currently analyzing survey data covering nearly 4 times the area
 used in the Y1 analyses of this paper and \citet{keypaper}, and there
 are ongoing improvements in depth,
calibration, and methodology.  
Thus we expect $>2\times$ reduction in
the statistical and systematic uncertainties in future cosmological
constraints, compared to the Y1 work.  Uncertainties in $n^i(z)$ will
become the dominant source of error in future analyses of DES and other imaging
surveys, without substantial improvement in the methodology presented here. 
We expect that the linear-shift approximation in \eqq{Dz}
will no longer suffice for quantifying the validation of the
$n^i(z)$.  Significant improvement will be needed in some combination of:
spectroscopic and/or multiband photometric validation data; photo-$z$
methodology; redshift range, bias constraints, and statistical errors
of WZ measurements; and treatment of survey inhomogeneity.  The
redshift characterization of broadband imaging surveys is 
a critical and active area of research, and will remain so in the
years to come.

\section*{Acknowledgements}

Support for DG was provided by NASA through Einstein Postdoctoral Fellowship grant
number PF5-160138 awarded by the Chandra X-ray Center, which is
operated by the Smithsonian Astrophysical Observatory for NASA under
contract NAS8-03060.

Funding for the DES Projects has been provided by the U.S. Department of Energy, the U.S. National Science Foundation, the Ministry of Science and Education of Spain, 
the Science and Technology Facilities Council of the United Kingdom, the Higher Education Funding Council for England, the National Center for Supercomputing 
Applications at the University of Illinois at Urbana-Champaign, the Kavli Institute of Cosmological Physics at the University of Chicago, 
the Center for Cosmology and Astro-Particle Physics at the Ohio State University,
the Mitchell Institute for Fundamental Physics and Astronomy at Texas A\&M University, Financiadora de Estudos e Projetos, 
Funda{\c c}{\~a}o Carlos Chagas Filho de Amparo {\`a} Pesquisa do Estado do Rio de Janeiro, Conselho Nacional de Desenvolvimento Cient{\'i}fico e Tecnol{\'o}gico and 
the Minist{\'e}rio da Ci{\^e}ncia, Tecnologia e Inova{\c c}{\~a}o, the Deutsche Forschungsgemeinschaft and the Collaborating Institutions in the Dark Energy Survey. 

The Collaborating Institutions are Argonne National Laboratory, the University of California at Santa Cruz, the University of Cambridge, Centro de Investigaciones Energ{\'e}ticas, 
Medioambientales y Tecnol{\'o}gicas-Madrid, the University of Chicago, University College London, the DES-Brazil Consortium, the University of Edinburgh, 
the Eidgen{\"o}ssische Technische Hochschule (ETH) Z{\"u}rich, 
Fermi National Accelerator Laboratory, the University of Illinois at Urbana-Champaign, the Institut de Ci{\`e}ncies de l'Espai (IEEC/CSIC), 
the Institut de F{\'i}sica d'Altes Energies, Lawrence Berkeley National Laboratory, the Ludwig-Maximilians Universit{\"a}t M{\"u}nchen and the associated Excellence Cluster Universe, 
the University of Michigan, the National Optical Astronomy Observatory, the University of Nottingham, The Ohio State University, the University of Pennsylvania, the University of Portsmouth, 
SLAC National Accelerator Laboratory, Stanford University, the University of Sussex, Texas A\&M University, and the OzDES Membership Consortium.

Based in part on observations at Cerro Tololo Inter-American Observatory, National Optical Astronomy Observatory, which is operated by the Association of 
Universities for Research in Astronomy (AURA) under a cooperative agreement with the National Science Foundation.

The DES data management system is supported by the National Science Foundation under Grant Numbers AST-1138766 and AST-1536171.
The DES participants from Spanish institutions are partially supported by MINECO under grants AYA2015-71825, ESP2015-88861, FPA2015-68048, SEV-2012-0234, SEV-2016-0597, and MDM-2015-0509, 
some of which include ERDF funds from the European Union. IFAE is partially funded by the CERCA program of the Generalitat de Catalunya.
Research leading to these results has received funding from the European Research
Council under the European Union's Seventh Framework Program (FP7/2007-2013) including ERC grant agreements 240672, 291329, and 306478.
We  acknowledge support from the Australian Research Council Centre of Excellence for All-sky Astrophysics (CAASTRO), through project number CE110001020.

This manuscript has been authored by Fermi Research Alliance, LLC under Contract No. DE-AC02-07CH11359 with the U.S. Department of Energy, Office of Science, Office of High Energy Physics. The United States Government retains and the publisher, by accepting the article for publication, acknowledges that the United States Government retains a non-exclusive, paid-up, irrevocable, world-wide license to publish or reproduce the published form of this manuscript, or allow others to do so, for United States Government purposes.

Based in part on zCOSMOS observations carried out using the Very Large Telescope at the ESO Paranal Observatory under Programme ID: LP175.A-0839.

\bibliographystyle{mn2e} 
\bibliography{photoz,des_y1kp_short}

\appendix
\section{Compensation for unknown covariances \magenta{in tomographic analyses}}
\label{correlationappendix}
There are four parameters $\Delta x_i$ in our model for errors in the
redshift distribution, which will be used when constraining some
parameter(s) $\pi$ of the cosmological model.  While we have produced
reliable bounds $\sigma^2_i$ of the variance of each of these, we have
less knowledge of the off-diagonal elements $r_{ij}\sigma_i\sigma_j$
of the covariance matrix $C$ of the $\Delta x_i$---perhaps we know
only that $|r_{ij}|\le r$.  We wish to make estimates of $\pi$ and the
uncertainty $\sigma_\pi$ that we are sure do not underestimate the
true error, for any allowed values of the $r_{ij}$, \magenta{in analyses that combine these redshift bins.}  We show here this
can be done by amplifying the diagonal elements of $C$ by a factor
$f^2$ while setting the off-diagonal elements to zero \magenta{(cf.~also~\citealt{shearcat}, their appendix D)}.

We consider a general case where a parameter $\pi$ depends on a vector
${\bf x}$ of $N$ elements via a linear relation $\pi = {\bf w}^T {\bf
  x}$ for some unit vector ${\bf w}$.  
Without loss of generality we can assume that the covariance matrix
$C$ of ${\bf x}$ has $C_{ii}=1$ and
$C_{ij}=r_{ij}$ for $i\ne j$.  Since $C$ is positive-definite,
$|r_{ij}|<1.$  Our task is to seek a value $f$ such that we can
guarantee that our estimate of the error on $\pi$ exceeds its true
uncertainty: 
\begin{equation}
{\bf w}^T (f^2 I) {\bf w} \ge \sigma^2_\pi = {\bf w}^T C {\bf w}, 
\end{equation}
for all unit vectors ${\bf w}$ and any $r_{ij}$ meeting our criteria.
Another way to view this is that we wish to construct a spherical
error region in ${\bf x}$ that is at least as large as the ellipsoid
defined by $C$ in every direction.

Clearly the condition is satisfied if and only if we can guarantee
that
\begin{equation}
f^2 \ge \lambda_{\rm max},
\end{equation}
where $\lambda_{\rm max}$ is the largest of the (positive) eigenvalues
$\lambda_i$ of $C$  ($i=1,\ldots,N$).  The eigenvalues are the
solutions of a polynomial equation
\begin{align}
0 & = | C - \lambda I| \\
 & = (1-\lambda)^N - (1-\lambda)^{N-2}\sum_{i>j} r^2_{ij} \\
 & \; \; \; \; + \left[\textrm{lower-order terms in}\; (1-\lambda)\right] \\
 & = \lambda^N - N\lambda^{N-1} + \left[ \frac{N(N-1)}{2} - \sum_{i>j}
   r_{ij}^2 \right] \lambda^{N-2} \\
 & \; \; \; \; + \left[\textrm{lower-order terms in}\;
   \lambda\right].
\end{align}
One can see that the roots of this polynomial must satisfy
\begin{align}
\sum_i \lambda_i & = N, \\
\textrm{Var}(\lambda) & = \frac{2}{N} \sum_{i>j} r^2_{ij} \le (N-1) r^2.
\end{align}
It is also straightforward to show that the maximum eigenvalue must be
within a certain distance of the mean eigenvalue:
\begin{equation}
f^2 = \lambda_{\rm max} \le \frac{\sum_i \lambda_i}{N} + \sqrt{(N-1)
  \textrm{Var}(\lambda)} \le 1 + (N-1)r.
\label{fmax}
\end{equation}
If we only know that $r<1$, then we must increase the diagonal
elements of the covariance matrix by $f=\sqrt{N}.$  This applies to
the case when all $N$ values of $x$ are fully correlated ($r=1$), and our
parameter responds to the mean of ${\bf x}.$

In the case of our $\Delta z^i$, we have $N=4,$ and we estimate that
correlation coefficients between bins should be modest, $|r_{ij}|\le
0.5$  (see Figure~\ref{fig:buzzardcorrelation}).  Then \eqq{fmax}
implies that inflating the individual bins' errors by
$f=\sqrt{2.5}\approx1.6$ will yield a conservative estimate of the
impact of redshift uncertainties on any parameter $\pi$.

\section{Effect of $n^i(z)$ inhomogeneities}
\label{inhomogeneity}
The DES Y1 analyses assume that the WL source galaxies in bin $i$ have
a redshift distribution $n^i(z)$ that is independent of sky position
$\theta,$ apart from the intrinsic density fluctuations in the
Universe.  Our survey is inhomogeneous in exposure time and seeing, however, and
furthermore is not properly corrected for Galactic extinction.
This induces angular fluctuations both in the overall source density $n_S$
and in the redshift distribution $n_S(z)$ of the galaxies in the bin
(here dropping the bin index $i$ for simplicity).  For a fixed lens
redshift, a fluctuation in source redshift distribution changes the mean
inverse critical density.  This produces a multiplicative deviation
between the measured shear and the true shear in some angular region,
which we will adopt as a rough description of the effect on shear
measurements even though the lenses are distributed in redshift:
\begin{equation}
\hat \gamma(\theta) = \left[1 + \epsilon(\theta)\right]\gamma(\theta). 
\end{equation}
We can similarly define a deviation of the mean source and lens
densities as
\begin{align}
n_L(\theta) & = \bar n_L \left[1 + \delta_L(\theta)\right], \\
n_S(\theta) & = \bar n_S \left[1 + \delta_S(\theta)\right], \\
\langle \delta_L \rangle & = \langle \delta_S \rangle = 0.
\label{deltals}
\end{align}
The averages above are over angular position $\theta$ within the footprint.
In the DES Y1 analyses, the lenses are \redmagic galaxies, which are
selected to be volume-limited and hence nominally have $\delta_L=0.$
To ensure that this is true, \citet{wthetapaper} look for any
correlation between $n_L(\theta)$ and observing conditions.  If any
such correlations are found, the lenses are reweighted to homogenize
the mean density.  We can assume therefore that $\delta_L=0$
everywhere, i.e. any fluctuations in lens density are much smaller
than those in the sources.

Both the determination of the shear response calibration \citep{shearcat}
and the validation of the redshift distribution (in this paper) are
produced with per-galaxy weighting, which means that the nominal shear
response is calibrated such that
\begin{align}
\frac{\left \langle n_S (1 + \epsilon) \right\rangle}{\langle n_S
  \rangle} & = 1 \\
\Rightarrow\quad \left\langle (1+\delta_S) \epsilon \right\rangle & = 0.
\label{nsepsilon}
\end{align}
We also assume that the source density and depth fluctuations are
uncorrelated with the shear signal, $\langle \delta \gamma \rangle =
\langle \epsilon \gamma \rangle = 0,$ since $\gamma$ is extragalactic
in origin while $\delta$ and $\epsilon$ have terrestrial or Galactic causes.

First we consider the galaxy-galaxy lensing observable \citep{gglpaper}.
It is an average of tangential shear of source galaxies about the
positions of lens galaxies.  Since it is calculated by summing over
lens-source pairs, the resultant measurement converges to
\begin{align}
\left\langle \hat\gamma_t(\theta) \right\rangle & = \frac{ \left\langle n_L
              n_S (1+\epsilon) \gamma_t(\theta) \right\rangle_\theta}
  {\left\langle  n_L  n_S \right\rangle_\theta}
  \\
 & = \left\langle (1+\delta_s) (1+\epsilon)
   \right\rangle_\theta \gamma_t(\theta) \\
 & = \gamma_t(\theta).
\end{align}
Here $\theta$ is the separation between lens and source, and the
averages are taken over lens-source pairs with separation in 
some range about $\theta.$  The last two lines are simplifications
that arise from $\delta_L=0$ and the vanishing conditions in
(\ref{deltals}) and (\ref{nsepsilon}) above.  The tangential-shear
measurement is, therefore, unaffected by survey inhomogeneity, as long
as the nominal shear and redshift calibrations are weighted by number
of source galaxies, not by area.

The other DES Y1 cosmological observable using the source population
is the two-point correlation function of shear $\xi_\gamma(\theta)$.
The shear $\gamma$ is a two-component field, and there are two
non-trivial correlation functions $\xi_\pm,$ or equivalently the spin
field can be decomposed into $E$ and $B$-mode components.
\citet{Guzik05} analyze the influence of multiplicative inhomogeneities
on the full $E/B$ field, and demonstrate that such systematic errors
shift power between $E$ and $B$ modes at a level comparable to the
change in the $E$ mode.  Here we will consider a simplified scalar
version of the \citet{Guzik05} formalism, which we can think of as
quantifying the $E$-mode errors due to inhomogeneity.  If these are
small, we do not have to worry about effects on $B$ modes either.

The calculation of $\xi_\gamma$ in \citet{shearcorr} accumulates the shear
products of all pairs of source galaxies 1 and 2 separated by angles in a
range near $\theta,$ yielding an estimate
\begin{align}
\hat \xi_\gamma(\theta) & = \frac{ \left\langle n_1 n_2
                   (1+\epsilon_1)(1+\epsilon_2) \gamma_1 \gamma_2
                   \right\rangle_\theta}
{\left\langle n_1 n_2 \right\rangle_\theta} \\
 & = \frac{ \left\langle (1+\delta_1)(1+\delta_2)
                   (1+\epsilon_1)(1+\epsilon_2) \gamma_1 \gamma_2
                   \right\rangle_\theta}
{\left\langle (1+\delta_1)(1+\delta_2) \right\rangle_\theta} \\
& \approx \left\langle \gamma_1 \gamma_2 \right\rangle_\theta \left[1 + \langle \delta_1 \delta_2 \rangle_\theta \right]^{-1} \times \\
& \;\;\;\;\;  \Big[ 1 + \langle \delta_1 \delta_2 \rangle_\theta  \nonumber \\
& \;\;\;\;\;\; + \left\langle(1+\delta_1)\epsilon_1\right\rangle_\theta
+ \left\langle(1+\delta_2)\epsilon_2\right\rangle_\theta + \left\langle \delta_2\epsilon_1\right\rangle_\theta
+ \left\langle \delta_1\epsilon_2\right\rangle_\theta \nonumber \\
&\;\;\;\;\;\;  + \left\langle \epsilon_1\epsilon_2\right\rangle_\theta \Big] \nonumber \\
& \approx \xi_\gamma(\theta) \left[1+ 2 \xi_{\delta\epsilon}(\theta) +
  \xi_\epsilon(\theta)\right].
\end{align}
We have kept terms only to second order in $\delta$ and $\epsilon$.
We also exploited the lack of
correlation between true shear and the systematic errors.
We find, following
\citet{Guzik05}, that the systematics lead to a multiplicative error in
$\xi_\gamma(\theta)$ given by the correlation function $\xi_\epsilon(\theta)$
of the multiplicative systematic; there are additional terms from the
cross-correlation $\xi_{\delta\epsilon}$ of density and depth inhomogeneities, which we
expect to be of the same order.
Since $\xi_\epsilon(\theta) \le \langle \epsilon^2 \rangle$, the
fractional error in $\xi_\gamma$ is no larger than the \emph{square}
of the typical fluctuation in source catalog density or inverse
critical density.  

The RMS fluctuation in source
mean redshift induced by failure to apply the SLR
 \RR{adjustment}  (\S\ref{knownerrors}) is $\delta z \lesssim0.01$.  
 \RR{We estimate the effect of variations in survey depth by removing sources above an $i$ band MOF magnitude $m_{\rm lim}$ from the matched COSMOS sample. The derivative of $\langle z\rangle$ w.r.t. $m_{\rm lim}$ is below 0.05 mag$^{-1}$ in the relevant range of $m_{\rm lim}$ for all four source bins. At the variation of depth present in DES Y1 (0.25 mag RMS, \citealt{y1gold}), this leads to a RMS fluctuation in source mean redshift of $\delta z \lesssim0.012$. Jointly, the RMS due to both effects, which may be partly correlated, are $\delta z \lesssim0.02$ RMS.}

 We expect the
source density and inverse critical density
(i.e. $\delta$ and $\epsilon$) to scale no faster than linearly with the mean
redshift of the sample, and the lowest redshift bin has
\RR{$z\approx0.3$, so $\langle \epsilon^2\rangle \lesssim (\delta z /z)^2
\approx 0.004$.}  Thus we estimate an overall scaling error of the
$\xi_\gamma$ measurements at roughly this level.

The most accurately measured combination of 
cosmological parameters in DES Y1 data is
$S_8 = \sigma_8 (\Omega_m/0.3)^{0.5}$, which is determined to a
fractional accuracy of $\approx3.5\%$ \citep{keypaper}.  Since $\xi_+\propto
S_8^2$, roughly, its error due to uncorrected Galactic extinction is
estimated to be \RR{$\approx 8\times$} smaller than the uncertainty level
in the DES Y1 analyses.  


\section{Calibration of DNF $n^i(z)$}
\label{dnfresults}

\begin{figure}
\centering
\includegraphics[width=\linewidth]{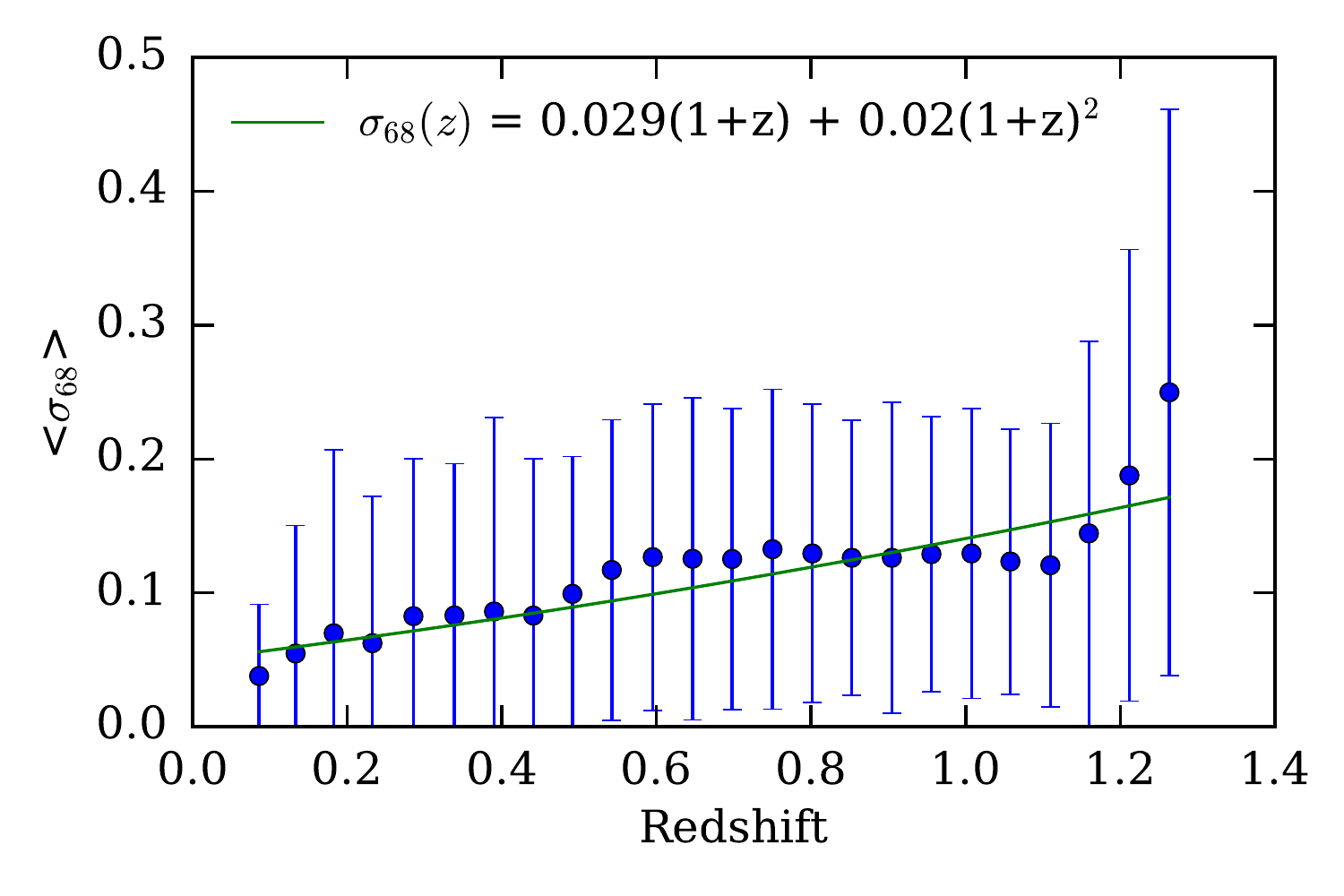}
\includegraphics[width=\linewidth]{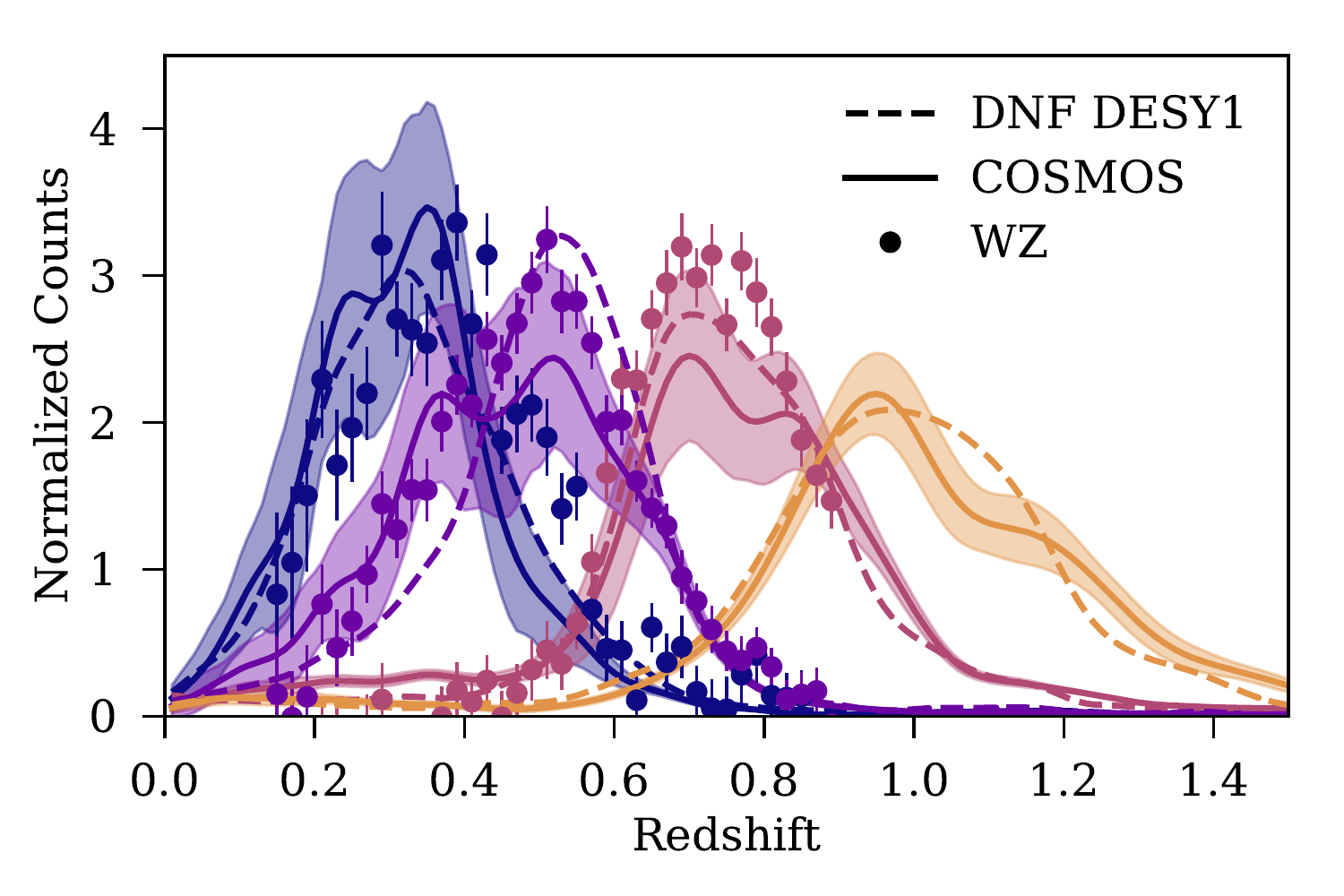}
\caption{Top panel: as Figure~\ref{sigma_bpz} showing the average width of the posterior distributions of DNF photometric redshifts. Bottom panel: the $n^i(z)$ for galaxies with bin assignments and estimated using DNF
  photo-$z$'s rather than BPZ. The error bars correspond to the standard deviation of the individual source's $\sigma_{68}$ around the average. The bottom panel is otherwise equivalent to
  Figure~\ref{fig:dndz}.} 
\label{dnfdndz}
\end{figure}

The COSMOS validation procedure of \S\ref{cosmosresample} was repeated
for the DNF photo-$z$'s in the same way as for BPZ, as was the WZ
validation.  The resultant $\Delta z^i$ are shown in
Table~\ref{dnftable} and the $n^i(z)$ and photo-$z$ precision metrics are plotted in
Figure~\ref{dnfdndz}.  Note that we do not require agreement between
the values for DNF and those for BPZ, because they apply to different
binnings of the source galaxies.

\begin{table*}
\caption{Values of and error contributions to photo-$z$ shift
  parameters of DNF $n^i(z)$}
\label{dnftable}
\begin{center}
\begin{tabular}{l|cccc|}
\hline
\hline
Value & Bin 1 & Bin 2 & Bin 3 & Bin 4  \\
\hline 
$z^{\rm PZ}$ range & 0.20--0.43 & 0.43--0.63 & 0.63--0.90 & 0.90-1.30\\[3pt]

COSMOS footprint sampling & 
$\pm0.0071$ &
$\pm0.0075$ &
$\pm0.0053$ &
$\pm0.0080$ \\

COSMOS limited sample size & 
$\pm0.0015$ &
$\pm0.0014$ &
$\pm0.0016$ &
$\pm0.0039$ \\

COSMOS photometric calibration errors & 
$\pm0.0023$ &
$\pm0.0029$ &
$\pm0.0046$ &
$\pm0.0045$ \\

COSMOS hidden variables &
$\pm0.0030$ &
$\pm0.0080$ &
$\pm0.0090$ &
$\pm0.0030$ \\

COSMOS errors in matching &
$\pm0.0069$ &
$\pm0.0069$ &
$\pm0.0069$ &
$\pm0.0069$ \\ [5pt]

COSMOS single-bin $\Delta z^i$ uncertainty & 
$\pm0.011$ & $\pm0.013$ & $\pm0.013$ & $\pm0.013$ \\[5pt]

\multicolumn{5}{c}{\metacal} \\[3pt]
 
COSMOS final $\Delta z^i$, \red{tomographic uncertainty} & 
$-0.024\pm0.017$ & 
$-0.042\pm0.021$ & 
$+0.006\pm0.021$ & 
$+0.038\pm0.020$ \\

WZ final $\Delta z^i$ & 
$+0.003\pm0.014$ & 
$-0.037\pm0.014$ & 
$+0.005\pm0.019$  & 
--- \\
\textbf{Combined final $\Delta z^i$} & 
$-0.008 \pm0.011$ &
$-0.039 \pm0.012$ &
$+0.006\pm0.014$ &
$+0.038 \pm0.020$ \\[5pt]
\hline
\end{tabular}
\end{center}
\end{table*}

\section*{Affiliations}

\scriptsize
$^{1}$ Universit\"ats-Sternwarte, Fakult\"at f\"ur Physik, Ludwig-Maximilians Universit\"at M\"unchen, Scheinerstr. 1, 81679 M\"unchen, Germany\\
$^{2}$ Max Planck Institute for Extraterrestrial Physics, Giessenbachstrasse, 85748 Garching, Germany\\
$^{3}$ Kavli Institute for Particle Astrophysics \& Cosmology, P. O. Box 2450, Stanford University, Stanford, CA 94305, USA\\
$^{4}$ SLAC National Accelerator Laboratory, Menlo Park, CA 94025, USA\\
$^{5}$ Department of Physics and Astronomy, University of Pennsylvania, Philadelphia, PA 19104, USA\\
$^{6}$ Centro de Investigaciones Energ\'eticas, Medioambientales y Tecnol\'ogicas (CIEMAT), Madrid, Spain\\
$^{7}$ Department of Physics \& Astronomy, University College London, Gower Street, London, WC1E 6BT, UK\\
$^{8}$ Department of Physics, ETH Zurich, Wolfgang-Pauli-Strasse 16, CH-8093 Zurich, Switzerland\\
$^{9}$ Institute of Space Sciences, IEEC-CSIC, Campus UAB, Carrer de Can Magrans, s/n,  08193 Barcelona, Spain\\
$^{10}$ Department of Physics, Stanford University, 382 Via Pueblo Mall, Stanford, CA 94305, USA\\
$^{11}$ Center for Cosmology and Astro-Particle Physics, The Ohio State University, Columbus, OH 43210, USA\\
$^{12}$ Department of Physics, The Ohio State University, Columbus, OH 43210, USA\\
$^{13}$ Institut de F\'{\i}sica d'Altes Energies (IFAE), The Barcelona Institute of Science and Technology, Campus UAB, 08193 Bellaterra (Barcelona) Spain\\
$^{14}$ Brookhaven National Laboratory, Bldg 510, Upton, NY 11973, USA\\
$^{15}$ ARC Centre of Excellence for All-sky Astrophysics (CAASTRO)\\
$^{16}$ School of Mathematics and Physics, University of Queensland,  Brisbane, QLD 4072, Australia\\
$^{17}$ Laborat\'orio Interinstitucional de e-Astronomia - LIneA, Rua Gal. Jos\'e Cristino 77, Rio de Janeiro, RJ - 20921-400, Brazil\\
$^{18}$ Observat\'orio Nacional, Rua Gal. Jos\'e Cristino 77, Rio de Janeiro, RJ - 20921-400, Brazil\\
$^{19}$ INAF - Osservatorio Astrofisico di Torino, Pino Torinese, Italy\\
$^{20}$ Department of Astronomy, University of Illinois, 1002 W. Green Street, Urbana, IL 61801, USA\\
$^{21}$ National Center for Supercomputing Applications, 1205 West Clark St., Urbana, IL 61801, USA\\
$^{22}$ Kavli Institute for Cosmological Physics, University of Chicago, Chicago, IL 60637, USA\\
$^{23}$ School of Physics and Astronomy, University of Southampton,  Southampton, SO17 1BJ, UK\\
$^{24}$ Fermi National Accelerator Laboratory, P. O. Box 500, Batavia, IL 60510, USA\\
$^{25}$ Centre for Astrophysics \& Supercomputing, Swinburne University of Technology, Victoria 3122, Australia\\
$^{26}$ Lawrence Berkeley National Laboratory, 1 Cyclotron Road, Berkeley, CA 94720, USA\\
$^{27}$ Australian Astronomical Observatory, North Ryde, NSW 2113, Australia\\
$^{28}$ Sydney Institute for Astronomy, School of Physics, A28, The University of Sydney, NSW 2006, Australia\\
$^{29}$ Department of Astronomy, The Ohio State University, Columbus, OH 43210, USA\\
$^{30}$ The Research School of Astronomy and Astrophysics, Australian National University, ACT 2601, Australia\\
$^{31}$ Institute of Cosmology \& Gravitation, University of Portsmouth, Portsmouth, PO1 3FX, UK\\
$^{32}$ Jodrell Bank Center for Astrophysics, School of Physics and Astronomy, University of Manchester, Oxford Road, Manchester, M13 9PL, UK\\
$^{33}$ Department of Physics, University of Arizona, Tucson, AZ 85721, USA\\
$^{34}$ Purple Mountain Observatory, Chinese Academy of Sciences, Nanjing, Jiangshu 210008, China\\
$^{35}$ Cerro Tololo Inter-American Observatory, National Optical Astronomy Observatory, Casilla 603, La Serena, Chile\\
$^{36}$ Department of Physics and Electronics, Rhodes University, PO Box 94, Grahamstown, 6140, South Africa\\
$^{37}$ LSST, 933 North Cherry Avenue, Tucson, AZ 85721, USA\\
$^{38}$ CNRS, UMR 7095, Institut d'Astrophysique de Paris, F-75014, Paris, France\\
$^{39}$ Sorbonne Universit\'es, UPMC Univ Paris 06, UMR 7095, Institut d'Astrophysique de Paris, F-75014, Paris, France\\
$^{40}$ George P. and Cynthia Woods Mitchell Institute for Fundamental Physics and Astronomy, and Department of Physics and Astronomy, Texas A\&M University, College Station, TX 77843,  USA\\
$^{41}$ Department of Physics, IIT Hyderabad, Kandi, Telangana 502285, India\\
$^{42}$ Department of Physics, California Institute of Technology, Pasadena, CA 91125, USA\\
$^{43}$ Jet Propulsion Laboratory, California Institute of Technology, 4800 Oak Grove Dr., Pasadena, CA 91109, USA\\
$^{44}$ Department of Astronomy, University of Michigan, Ann Arbor, MI 48109, USA\\
$^{45}$ Department of Physics, University of Michigan, Ann Arbor, MI 48109, USA\\
$^{46}$ Instituto de Fisica Teorica UAM/CSIC, Universidad Autonoma de Madrid, 28049 Madrid, Spain\\
$^{47}$ Institute of Astronomy, University of Cambridge, Madingley Road, Cambridge CB3 0HA, UK\\
$^{48}$ Kavli Institute for Cosmology, University of Cambridge, Madingley Road, Cambridge CB3 0HA, UK\\
$^{49}$ Department of Astronomy, University of California, Berkeley,  501 Campbell Hall, Berkeley, CA 94720, USA\\
$^{50}$ Astronomy Department, University of Washington, Box 351580, Seattle, WA 98195, USA\\
$^{51}$ Santa Cruz Institute for Particle Physics, Santa Cruz, CA 95064, USA\\
$^{52}$ Argonne National Laboratory, 9700 South Cass Avenue, Lemont, IL 60439, USA\\
$^{53}$ Departamento de F\'isica Matem\'atica, Instituto de F\'isica, Universidade de S\~ao Paulo, CP 66318, S\~ao Paulo, SP, 05314-970, Brazil\\
$^{54}$ Department of Astrophysical Sciences, Princeton University, Peyton Hall, Princeton, NJ 08544, USA\\
$^{55}$ Instituci\'o Catalana de Recerca i Estudis Avan\c{c}ats, E-08010 Barcelona, Spain\\
$^{56}$ Department of Physics and Astronomy, Pevensey Building, University of Sussex, Brighton, BN1 9QH, UK\\
$^{57}$ Instituto de F\'\i sica, UFRGS, Caixa Postal 15051, Porto Alegre, RS - 91501-970, Brazil\\
$^{58}$ Instituto de F\'isica Gleb Wataghin, Universidade Estadual de Campinas, 13083-859, Campinas, SP, Brazil\\
$^{59}$ Computer Science and Mathematics Division, Oak Ridge National Laboratory, Oak Ridge, TN 37831\\
$^{60}$ Excellence Cluster Universe, Boltzmannstr.\ 2, 85748 Garching, Germany\\
$^{61}$ Institute for Astronomy, University of Edinburgh, Edinburgh EH9 3HJ, UK
\end{document}